# Experimental verification of field-enhanced molecular vibrational scattering at single infrared antennas

**Divya Virmani,[1] Carlos Maciel-Escudero,[1,2] Rainer Hillenbrand,[1,3,4]\* and Martin Schnell[1,3]\***

*[1] CIC nanoGUNE BRTA, 20018 Donostia-San Sebastián, Basque Country, Spain*

*[2] Materials Physics Center, CSIC-UPV/EHU, 20018 Donostia-San Sebastián, Spain*

*[3] IKERBASQUE, Basque Foundation for Science, 48013 Bilbao, Basque Country, Spain*

*[4] Department of Electricity and Electronics, UPV/EHU, 20018 Donostia-San Sebastián, Spain*

*\*r.hillenbrand@nanogune.eu*

*\*schnelloptics@gmail.com*

**Abstract**

Surface-enhanced infrared absorption (SEIRA) spectroscopy exploits the field enhancement near nanophotonic structures for highly sensitive characterization of (bio)molecules. The vibrational signature observed in SEIRA spectra is typically interpreted as field-enhanced molecular absorption. Here we study molecular vibrations in the near field of single antennas and show that the vibrational signature can be equally well explained by field-enhanced molecular scattering. Although the infrared scattering cross section of molecules is negligible compared to their absorption cross section, the interference between the molecular-scattered field and the incident field enhances the spectral signature caused by molecular vibrational scattering by 10 orders of magnitude, thus becoming as large as that of field-enhanced molecular absorption. We provide experimental evidence that field-enhanced molecular scattering can be measured, scales in intensity with the fourth power of the local field enhancement and fully explains the vibrational signature in SEIRA spectra in both magnitude and line shape. Our work may open new paths for developing highly sensitive SEIRA sensors that exploit the presented scattering concept.

**Introduction**

Infrared (IR) spectroscopy is a widely applied technique for the non-destructive and label-free analysis of materials based on their characteristic vibrational spectra[1]. Owing to the small absorption cross section of molecular vibrations, standard IR spectroscopy typically requires a large quantity of analytes to obtain a measurable signal. This limitation can be overcome by using surface-enhanced IR absorption (SEIRA) spectroscopy, where nanostructures provide strongly confined and enhanced fields at the substrate surface[2,3]. When molecules are placed in a region of strong field enhancement, the spectral signature of molecular vibrations is significantly increased and infrared spectroscopy of very small amounts of analytes becomes possible. Very high sensitivity can be achieved by exploiting nanostructures with pronounced resonances including metal[4–6] and graphene[7] nanoantennas (localized surface plasmon resonance), van-der-Waals material nanostructures (phonon resonances)[8] and dielectric nanoresonators[9,10]. These SEIRA architectures allow for increasing the spectral signature by at least five orders of magnitude compared to standard IR spectroscopy, which has helped to push the sensitivity of IR spectroscopy down to the attogram scale.

The enhancement mechanism in SEIRA is generally described as the coupling of molecular vibrations with the antenna. The resonant character of this coupling leads to asymmetric line shapes of molecular vibrations, which is a typical signature of SEIRA spectra and challenges the direct extraction of molecular information. A variety of theoretical models have been developed to fit and provide an intuitive understanding of the observed line shapes[2,11–15]. Often, vibrational line shapes in SEIRA are described as the Fano-like interference between the electromagnetic field of the (spectrally broad) antenna resonance and the field associated with the (spectrally narrow) molecular vibrations. Other models interpret SEIRA in the picture of coupled resonators and describe the vibrational line shapes as a coupled dark-bright mode system based on the effect of intrinsic and external losses[11], or in the picture of two coupled dipoles to predict vibrational line shapes quantitatively and explore the impact of the local field enhancement on the line shapes[15].

So far, it is widely accepted that field-enhanced molecular absorption is the underlying process for the appearance of vibrational lines in SEIRA spectra. This idea is rooted in the fact that the molecular absorption cross section is many orders of magnitude larger than the molecular scattering cross section at infrared frequencies. Further, experimental observations show that the spectral signature of molecular vibrations scales with the second power of the local field enhancement, $|f|^2$.[16] On the other hand, Alonso-Gonzalez et al. verified experimentally that the infrared elastic light scattering from a small non-absorbing particle in the vicinity of a single antenna is analogous to the mechanism of surface-enhanced Raman spectroscopy (SERS), that is, the intensity of the infrared elastic light scattering from the particle is enhanced by the fourth power of the local field enhancement, $|f|^4$.[17] The underlying process was described as scattering from the particle via the antenna after being illuminated by the antenna.[18,19] Rezus and Selig then assumed a small particle exhibiting molecular vibrations in the vicinity of the antenna and showed in a numerical study that field-enhanced molecular scattering fully describes the line shapes in extinction spectra of the antenna-particle system.[15] Experimental verification of field-enhanced molecular scattering as well as its connection with the vibrational line shapes is, however, challenging because of the difficulty to measure the scattered light from the antenna free of any interference with the incident field.

Here we explore the role of molecular scattering when mid-infrared molecular vibrations are located in the vicinity of a single antenna – a system that is usually considered in SEIRA experiments. We developed and numerically verified an analytical model that describes the vibrational line shapes in the extinction cross section as the interference between field-enhanced molecular scattering and the incident field. We then performed a near-field experiment that provides evidence that field-enhanced molecular scattering can be measured, scales in intensity with the fourth power of the field enhancement provided by the antenna, $|f|^4$, and fully explains the vibrational line shapes in the extinction cross section.

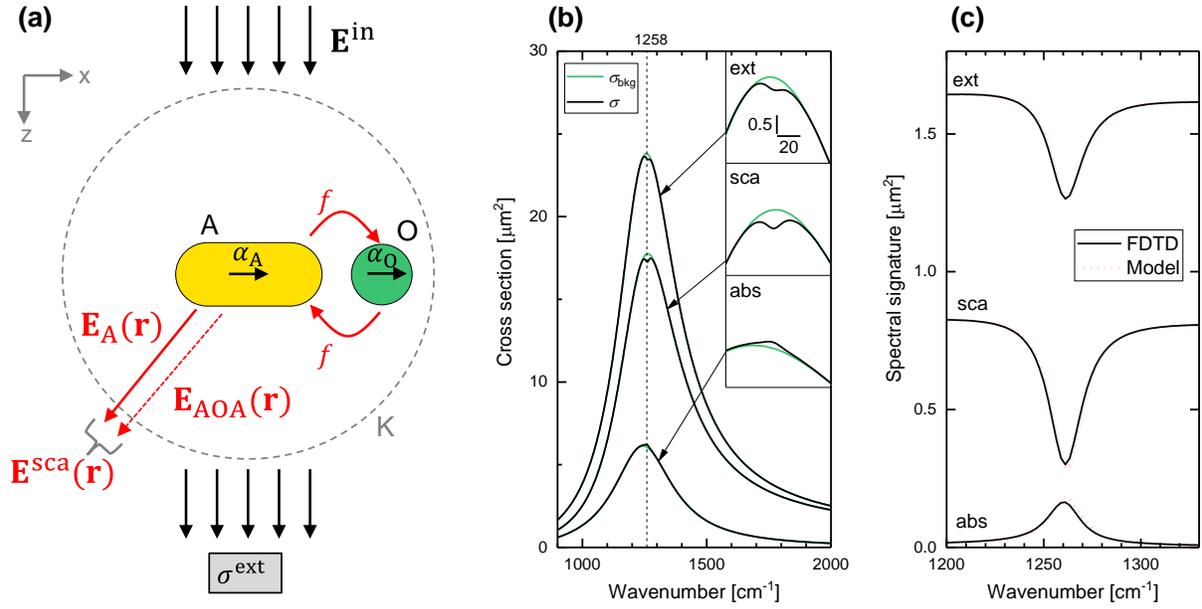

**Figure 1: Surface-enhanced infrared vibrational spectroscopy modeled as a scattering process.** (a) Elastic light scattering of a small particle exhibiting molecular vibrations (O) in the presence of an infrared-resonant antenna (A). (b) Numerically calculated extinction (ext), scattering (sca) and absorption (abs) cross section of the system in (a), where the antenna resonance is tuned to the vibrational resonance of the small particle at 1,258 cm$^{-1}$ ($\sigma$, black line). For reference, the case of a non-absorbing small particle is shown ($\sigma_{bkg}$, green line). (c) Baseline-corrected spectra, as obtained from (b), $\Delta\sigma = \sigma - \sigma_{bkg}$, show the spectral signature of the molecular vibrations (black line), which is fully explained by the scattering model in (a) ($\sigma_{vib}^{ext}$, Eq. (13), red dotted line). $\mathbf{E}^{in}$: Incident field. $\mathbf{E}_A$: Direct antenna scattering. $\mathbf{E}_{AOA}$: Field-enhanced molecular scattering. $\mathbf{E}^{sca}$: Total scattered field. $\sigma^{ext}$: Detector measuring light extinction. $\alpha_A$ and $\alpha_O$: point polarizabilities describing the antenna and small particle, respectively. $f$: field enhancement provided by the antenna. K: Sphere surrounding the antenna and the small particle.

**Scattering model**

In Fig. 1(a), we describe the typical SEIRA configuration where a small spherical object (O) exhibiting molecular vibrations is located in the near field of a single antenna (A).[15,17] The antenna-object system is illuminated by a monochromatic propagating plane wave,

$$\mathbf{E}^{in}(\mathbf{r}) = \mathbf{E}_0^{in} \exp(i\mathbf{k} \cdot \mathbf{r}), \qquad (1)$$

where $\mathbf{E}_0^{in} = E_0^{in} \hat{\mathbf{x}}$ is a constant vector describing the amplitude and polarization (x-direction) of the plane wave, which propagates along the direction $\hat{\mathbf{n}} = \mathbf{k}/k$ (z-direction) with a wavenumber $k$. The scattered field of the antenna-object system $\mathbf{E}^{sca}$ may be described as an interaction series that is truncated in the limit of a weakly scattering object:[18,19]

$$\mathbf{E}^{sca}(\mathbf{r}) = \mathbf{E}_A(\mathbf{r}) + \mathbf{E}_O(\mathbf{r}) + \mathbf{E}_{OA}(\mathbf{r}) + \mathbf{E}_{AO}(\mathbf{r}) + \mathbf{E}_{AOA}(\mathbf{r}) + \cdots. \tag{2}$$

In the case of elastic light scattering of a small object in the infrared spectral range, the scattered field $\mathbf{E}^{sca}(\mathbf{r})$ and the scattered field intensity $|\mathbf{E}^{sca}(\mathbf{r})|^2$ can be approximated by the following expressions for large field enhancement factors, $|f| \gg 1$, [17]:

$$\mathbf{E}^{sca}(\mathbf{r}) \approx \mathbf{E}_A(\mathbf{r}) + \mathbf{E}_{AOA}(\mathbf{r}), \tag{3}$$

$$|\mathbf{E}^{sca}(\mathbf{r})|^2 \approx \mathbf{E}_A^*(\mathbf{r}) \cdot \mathbf{E}_A(\mathbf{r}) + 2\mathrm{Re}\{\mathbf{E}_A^*(\mathbf{r}) \cdot \mathbf{E}_{AOA}(\mathbf{r})\}. \tag{4}$$

For simplicity, antenna (A) and object (O) are approximated as point dipoles with polarizability, $\alpha_A$ and $\alpha_O$, so that

$$\mathbf{E}_A(\mathbf{r}) = k^2 \mathbf{G}_x(\mathbf{r}, \mathbf{r}_A) \alpha_A E_x^{in}(\mathbf{r}_A), \tag{5}$$

$$\mathbf{E}_{AOA}(\mathbf{r}) = k^2 \mathbf{G}_x(\mathbf{r}, \mathbf{r}_A) f \alpha_O f E_x^{in}(\mathbf{r}_A), \tag{6}$$

where $\mathbf{G}_x(\mathbf{r}, \mathbf{r}_A)$ is the x-component of the free-space Green's tensor function and $\mathbf{r}_A$ is the position of the antenna. $f$ is the local field enhancement provided by the antenna at the position of the object and it is important to note that $f$ is generally a complex quantity, i.e. $f = |f|e^{i\mathrm{Arg}\{f\}}$, because of the resonant character of the antenna response. The index A indicates direct antenna scattering of the incident field $\mathbf{E}^{in}$. The index AOA refers to the scattering path of a double scattering process between antenna and object, where the incident field $\mathbf{E}^{in}$ excites the antenna, the antenna polarizes the object and the object scatters via the antenna to the far field. The AOA term carries information on the molecular vibration and is termed the *field-enhanced molecular scattering*. Note that terms to second or higher order in O can be neglected owing to the assumption of small object polarizability, $\alpha_O$. Having obtained an expression for the scattered field $\mathbf{E}^{sca}$, the extinction cross section may be derived by applying the optical theorem,[20]

$$\sigma^{ext} = \frac{4\pi}{k|\mathbf{E}_0^{in}|^2} \mathrm{Im}\left\{\mathbf{E}_0^{in*} \cdot \mathbf{E}_0^{sca}(\hat{\mathbf{n}})\right\}, \tag{7}$$

which relates the extinction cross section $\sigma^{ext}$ of the antenna-object system to the scattering amplitude $\mathbf{E}_0^{sca}(\hat{\mathbf{n}})$, defined as $\mathbf{E}^{sca}(\mathbf{r}) = \frac{\exp(ikr)}{r} \mathbf{E}_0^{sca}(\hat{\mathbf{r}})$ and evaluated in the propagation direction of the incident plane wave $\hat{\mathbf{n}}$.[21] The scattering cross section may be obtained by integration of the scattered field intensity, $|\mathbf{E}^{sca}|^2$, over the surface of a sphere K surrounding the antenna-object system[19,22]

$$\sigma^{sca} = \frac{1}{|\mathbf{E}_0^{in}|^2} \int_K dK\, |\mathbf{E}^{sca}(\mathbf{r})|^2. \tag{8}$$

With Eqs. (3),(4), the extinction and scattering cross sections in Eqs. (7) and (8) approximate to

$$\sigma^{\text{ext}} \approx \frac{4\pi}{k|\mathbf{E}_0^{\text{in}}|^2} \text{Im}\left\{\mathbf{E}_0^{\text{in}*} \cdot \mathbf{E}_{A,0}(\hat{\mathbf{n}}) + \mathbf{E}_0^{\text{in}*} \cdot \mathbf{E}_{AOA,0}(\hat{\mathbf{n}})\right\}, \qquad (9)$$

$$\sigma^{\text{sca}} \approx \frac{1}{|\mathbf{E}_0^{\text{in}}|^2} \int_K dK \, (\mathbf{E}_A^*(\mathbf{r}) \cdot \mathbf{E}_A(\mathbf{r}) + 2\text{Re}\{\mathbf{E}_A^*(\mathbf{r}) \cdot \mathbf{E}_{AOA}(\mathbf{r})\}), \qquad (10)$$

where $\mathbf{E}_{A,0}$ and $\mathbf{E}_{AOA,0}$ are the scattering amplitudes of the fields $\mathbf{E}_A(\mathbf{r})$ and $\mathbf{E}_{AOA}(\mathbf{r})$, respectively. Equation (9) provides a description of the extinction cross-section of the antenna-molecule system that is exclusively based on the incident field $\mathbf{E}^{\text{in}}$, the field associated with direct antenna scattering $\mathbf{E}_A$ and the field associated with field-enhanced molecular scattering $\mathbf{E}_{AOA}$. Molecular absorption does not explicitly appear as a term in Eq. (9), which is expected because the optical theorem is based on exclusively evaluating the scattered field. Specifically, the antenna resonance in the extinction cross section (Fig. 1(b), dashed line) is described by the interference of the incident field $\mathbf{E}^{\text{in}}$ with the direct antenna scattering $\mathbf{E}_A$ (the 1st term in Eq. (9)). The spectral signature of the molecular vibration in extinction cross section (Fig. 1(c)) is determined by the interference of the incident field $\mathbf{E}^{\text{in}}$ with the field-enhanced molecular scattering $\mathbf{E}_{AOA}$ (the 2nd term in Eq. (9)). In case of the scattering cross section (Eq. (10)), the spectral signature of the molecular vibration is caused by the interference of field-enhanced molecular scattering $\mathbf{E}_{AOA}$ with the direct antenna scattering $\mathbf{E}_A$ rather than the incident field (the 2nd term in Eq. (10)), explaining the differences in the magnitude and shape of the vibrational line shapes (Fig. 1(c), Extended Data Fig. 2). We obtain simple expressions for the extinction and scattering cross sections by inserting Eqs. (5),(6) in Eqs. (9),(10):

$$\sigma^{\text{ext}} = k\text{Im}\{\alpha_A + f^2\alpha_O\}, \qquad (11)$$

$$\sigma^{\text{sca}} = \frac{k^4}{6\pi} [|\alpha_A|^2 + 2\text{Re}\{\alpha_A^* f^2 \alpha_O\}]. \qquad (12)$$

We identify the 1st term in Eqs. (11), (12) as the extinction and scattering cross sections of the unloaded antenna, $\sigma_A^{\text{ext}} = k\text{Im}\{\alpha_A\}$ and $\sigma_A^{\text{sca}} = \frac{k^4}{6\pi}|\alpha_A|^2$. The 2nd term in Eqs. (11), (12) is the spectral signature of the molecular vibration in the extinction and scattering cross section,

$$\sigma_{\text{vib}}^{\text{ext}} = k\text{Im}\{f^2\alpha_O\}, \qquad \sigma_{\text{vib}}^{\text{sca}} = \frac{k^4}{3\pi}\text{Re}\{\alpha_A^* f^2 \alpha_O\}. \qquad (13)$$

Note that ref. 15 arrived at the same expression for $\sigma_{\text{vib}}^{\text{ext}}$ based on a model of two coupled point dipoles. Further note that the absorption cross section and the corresponding spectral signature of the molecular vibration can be directly obtained by applying energy conservation, $\sigma_{\text{vib}}^{\text{abs}} = \sigma_{\text{vib}}^{\text{ext}} - \sigma_{\text{vib}}^{\text{sca}}$. Equation (13) provides an intuitive explanation for the vibrational line shapes: the apparent polarizability of the object is increased by the square of the field enhancement $f$ provided by the antenna. Note that the field enhancement $f$ is generally a complex quantity and shifts the phase of the object polarizability $\alpha_O$. When taking the imaginary part in (Eq. (13)), the spectral signature in the extinction cross section $\sigma_{\text{vib}}^{\text{ext}}$ may thus yield different line shapes depending on the tuning of the antenna with respect to the molecular vibration.[15] In the case where the antenna resonance is tuned to the molecular vibration (that is, $\text{Arg}(f) = \pi/2$ and also $\text{Arg}(\alpha_A) = \pi/2$ because of $f = k^2 G_{xx}(\mathbf{r}_O, \mathbf{r}_A)\alpha_A$), Eq. (13) becomes

$$\sigma_{\text{vib}}^{\text{ext}} = -k|f|^2 \text{Im}\{\alpha_O\}, \qquad \sigma_{\text{vib}}^{\text{sca}} = -\frac{k^4}{3\pi}|\alpha_A||f|^2 \text{Im}\{\alpha_O\}. \tag{14}$$

At antenna resonance, the spectral signature of the molecular vibration in the extinction cross section, $\sigma_{\text{vib}}^{\text{ext}}$, based on an exclusively scattering model is formally identical with field-enhanced molecular absorption, $k|f|^2\text{Im}\{\alpha_O\}$, which is the typical accepted explanation of SEIRA. Thus, the vibrational signal in typical SEIRA spectra of single resonant antennas can be equivalently well understood as field-enhanced molecular absorption or interferometrically- and field-enhanced molecular scattering.

**Interferometric Enhancement**

It may seem counterintuitive to attribute the spectral signature of molecular vibrations in the extinction cross section to molecular scattering. Considering a spherical nanoparticle with a radius of 1 nm and a vibrational resonance at 1,258 cm$^{-1}$, the scattering cross section, $\frac{k^4}{6\pi}|\alpha_O|^2 \sim 2.3 \cdot 10^{-31}$ m$^2$ is ten orders of magnitude smaller than the absorption cross section, $k\text{Im}\{\alpha_O\} \sim 2.2 \cdot 10^{-21}$ m$^2$, and thus negligible (A comparison with other nanoparticle sizes is provided in Extended Data Table 1). Even the field enhancement provided by the antenna ($f \sim 50$) cannot compensate for this large discrepancy, that is, the field-enhanced molecular scattering cross section, $\frac{k^4}{6\pi}|f|^4|\alpha_O|^2 \sim 1.7 \cdot 10^{-24}$ m$^2$, is still six orders of magnitudes smaller than the field-enhanced molecular absorption cross section, $k|f|^2\text{Im}\{\alpha_O\} \sim 6.1 \cdot 10^{-18}$ m$^2$. We point out that these considerations apply to intensity measurements of pure field-enhanced molecular scattering and absorption. However, according to the presented model, the spectral signature is determined by the interference of field-enhanced molecular scattering $\mathbf{E}_{\text{AOA}}$ with the incident field $\mathbf{E}^{\text{in}}$ or the direct antenna scattering $\mathbf{E}_A$ (Eqs. (9),(10)), yielding an extraordinarily large enhancement of molecular scattering by a factor of

$$F^{\text{ext}} = \frac{|\sigma_{\text{vib}}^{\text{ext}}|}{\sigma_O^{\text{sca}}} \approx |f|^2 \frac{6\pi}{k^3|\alpha_O|} \approx 2.9 \cdot 10^{13}, \tag{15}$$

$$F^{\text{sca}} = \frac{|\sigma_{\text{vib}}^{\text{sca}}|}{\sigma_O^{\text{sca}}} \approx |f|^2 \frac{2|\alpha_A|}{|\alpha_O|} \approx 4.5 \cdot 10^{13}, \tag{16}$$

where $\sigma_{\text{vib}}^{\text{ext}}$ and $\sigma_{\text{vib}}^{\text{sca}}$ is the spectral signal of molecular vibrations in extinction and scattering cross section (Eq. (14)), respectively, compared to the molecular scattering cross section $\sigma_O^{\text{sca}} = \frac{k^4}{6\pi}|\alpha_O|^2$. In case of the extinction cross section (Eq. (15)), the interference of field-enhanced molecular scattering $\mathbf{E}_{\text{AOA}}$ with the incident field $\mathbf{E}^{\text{in}}$ provides an interferometric enhancement of $6\pi/k^3|\alpha_O| \approx 1.0 \cdot 10^{10}$. In case of the scattering cross section (Eq. (16)), the interference of field-enhanced molecular scattering $\mathbf{E}_{\text{AOA}}$ with the direct antenna scattering $\mathbf{E}_A$ provides an interferometric enhancement of $2|\alpha_A|/|\alpha_O| \sim 1.6 \cdot 10^{10}$. In this latter case, the dual role of the antenna becomes apparent in providing both field enhancement and interferometric enhancement of molecular scattering. The direct antenna scattering $\mathbf{E}_A$ can be interpreted as an internal reference field to $\mathbf{E}_{\text{AOA}}$ and may become very large compared to $\mathbf{E}_{\text{AOA}}$ owing to the much larger polarizability of the antenna, $\alpha_A$, compared to the object polarizability, $\alpha_O$. Together with the effect of the field enhancement ($|f|^2 \sim 2.7 \cdot 10^3$), interferometric enhancement provides a total enhancement of molecular scattering by 13 orders of magnitude, thus overcoming the large discrepancy between pure molecular scattering and absorption cross sections and providing an intuitive understanding of the enhancement mechanism in this scattering model.

**Numerical Validation**

To corroborate the presented scattering model, we performed numerical calculations of the configuration shown in Fig. 1(a) and compared the calculated spectral signature of the molecular vibration against the analytical expressions of $\sigma_{\text{vib}}^{\text{ext}}$ and $\sigma_{\text{vib}}^{\text{sca}}$ from Eq. (13) (Methods). In the case of an antenna tuned to the molecular vibration, the spectral signature of the molecular vibration is clearly recognizable in form of a dip and the scattering model (red line in Fig. 1(c)) fully reproduced the numerically calculated spectral signature, $\Delta\sigma^{\text{ext}}$ and $\Delta\sigma^{\text{sca}}$ (black line), in both magnitude and peak shape. The deviation was found to be 4.4% and 2.3% for the extinction and scattering cross section, respectively, which we attribute to numerical error (i.e. the challenge to accurately model the very small object (O) next to a large antenna). We further verified the scattering model for the following cases. First, the model yielded very good agreement for a range of object sizes (Extended Data Figure 1). Second, the model accurately reproduced the asymmetric (Fano-like) line shapes of SEIRA obtained with off-resonant antennas (Extended Data Figure 2). Briefly, antennas driven far below resonance (short antennas, $\text{Arg}(f) \to 0$) yield a spectral signature of the molecular vibration in form of a peak in the extinction cross section ($\text{Im}\{\alpha_O\}$). Antennas tuned to the molecular resonance ($\text{Arg}(f) = \pi/2$) yield a dip ($-\text{Im}\{\alpha_O\}$) and antennas driven far above resonance (long antennas, $\text{Arg}(f) \to \pi$) yield again a peak ($\text{Im}\{\alpha_O\}$). Because of the different interference process (cf. Eqs. (9), (10)), the vibrational line shape in the scattering cross section follows the sequence from normal dispersive ($\text{Re}\{\alpha_O\}$) to a dip ($-\text{Im}\{\alpha_O\}$) to anomalous dispersive ($-\text{Re}\{\alpha_O\}$) as the antenna length is varied from short to tuned to long. Third, the model reproduced accurately the coupling of the antenna with weak oscillators (molecular vibrations), but showed significant deviation in case of strong oscillators (phononic-like), which is expected because the assumption of weak coupling is built into the scattering model (Extended Data Figure 3).

**Experimental verification of field-enhanced molecular scattering**

To support the presented scattering model experimentally, and to provide evidence that the interaction between antenna and molecular vibration is indeed a scattering process, we designed an experiment that allowed us to directly measure the spectral signature of the field-enhanced molecular scattering $\mathbf{E}_{AOA}$ free of any interference with the incident field $\mathbf{E}^{in}$ or direct antenna scattering $\mathbf{E}_A$. To this end, we study the elastic light scattering from molecules located on a near-field probe, which is brought into the proximity of a metal nanorod antenna (Fig. 2). The molecules on the near-field probe mimic the molecules in a traditional SEIRA experiment, and by placing them on the near-field probe, we have the possibility of modulating the molecule-antenna distance, which in turn allows us to isolate the field-enhanced molecular scattering $\mathbf{E}_{AOA}$. In detail, we adopted and extended the approach from ref.[17] by using metalized atomic force microscopy (AFM) tips that are known to be contaminated by a thin layer of PDMS (Polydimethylsiloxane) as a proxy for object (O) with a vibrational resonance[23,24]. Both near-field probe and antenna were illuminated from below in a transflection geometry[25] with the weakly focused beam from a mid-infrared broadband laser. Illumination from below ensured efficient excitation of the antenna while avoiding direct excitation of the near-field probe. As in ref.[17], illumination and detection directions are the same, and as a consequence of reciprocity, it is thus ensured that the field enhancement of the illumination and scattering pathways are equal. The antenna scattered light was collected below the sample and detected interferometrically using a commercial nano-FTIR system (NeaSCOPE, attocube systems AG). This interferometric detection, combined with vertical vibration of the near-field probe at a frequency $\Omega$ and subsequent demodulation of the detector signal at a frequency $n\Omega$, yielded the *n*-th demodulation order, $\tilde{s}_n$, in amplitude $s_n$ and phase $\varphi_n$ (Methods):

$$\tilde{s}_n(\omega) = s_n(\omega)e^{i\varphi_n(\omega)} \propto E_{AOA}(\omega) \,. \tag{17}$$

The proportionality in Eq. (17) was shown in ref. [17] for the case of infrared-resonant rod antennas as considered here. For direct comparison, we express the field-enhanced molecular scattering $\mathbf{E}_{AOA}$ using the model in Eq. (6) in amplitude and phase:

$$|E_{AOA}| \propto k^2|f|^2|\alpha_O| \,, \tag{18}$$

$$\varphi_{AOA} = 2\text{Arg}\{f\} + \text{Arg}\{\alpha_O\}\,. \tag{19}$$

where field enhancement, $f$, was obtained numerically and particle polarizability, $\alpha_O$, was evaluated analytically (see Methods).

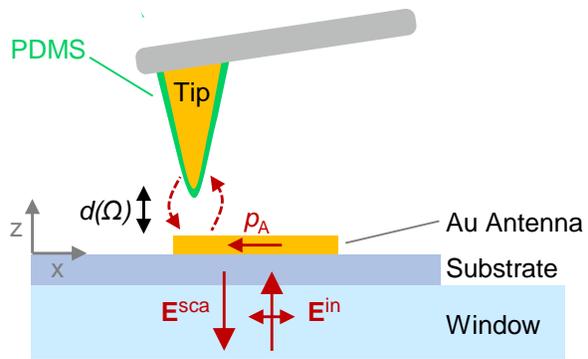

**Figure 2: Near-field experiment to measure field-enhanced molecular scattering.** $E^{\text{in}}$: Incident field polarized along the antenna long axis, $p_A$: Induced dipole moment in the antenna. $E^{\text{sca}}$: Scattered field, PDMS: Polydimethylsiloxane. Dashed lines illustrate the near-field coupling between antenna and tip, which is modulated by varying sinusoidally the tip-antenna distance, $d(\Omega)$, at frequency $\Omega$.

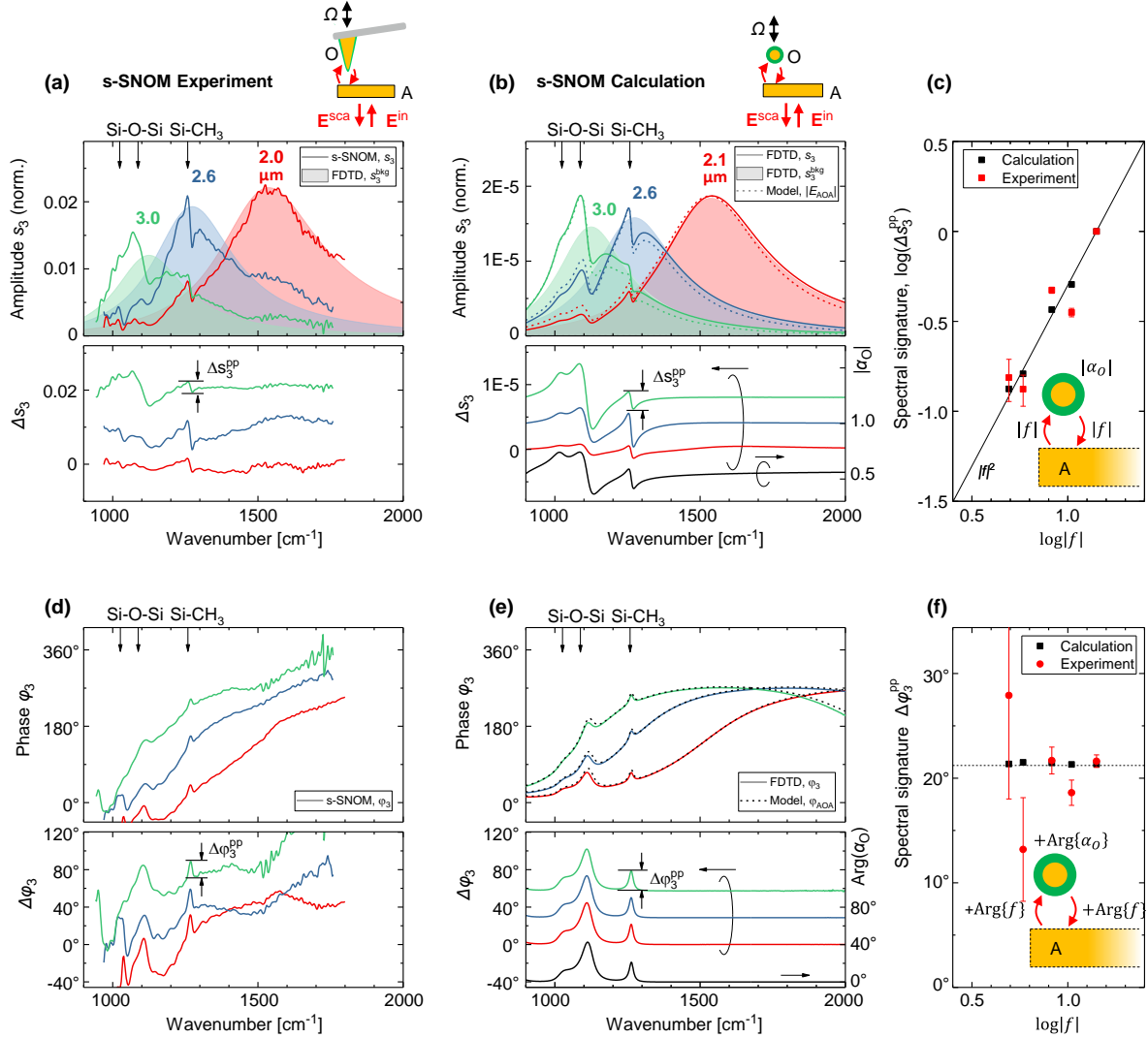

**Figure 3: Near-field spectroscopic measurement of field-enhanced molecular scattering.** (a) Experimental and (b) numerically calculated amplitude spectra, $s_3$, obtained with a PDMS-contaminated s-SNOM probe (O) that interacts with a metal nanorod (A) of different lengths, $L$ (solid lines). Shaded areas show the calculated near-field spectra, $s_3^{\text{bkg}}$, assuming a non-absorbing object. Bottom panel shows the isolated spectral signature of the molecular vibrations, $\Delta s_3 = s_3 - s_3^{\text{bkg}}$, (curves offset for clarity). The magnitude of the object polarizability, $|\alpha_O|$, is shown for reference (black line). For panel (b) only: Dashed lines show the magnitude of the AOA term, $|E_{\text{AOA}}|$, of the scattering model (Eq. 18). (c) Parametric representation of the spectral signature of the 1,258 cm$^{-1}$ (Si-CH$_3$) molecular vibration of PDMS, $\Delta s_3^{\text{pp}}$, as obtained from panel (a,b) (normalized to maximum), and plotted against the calculated field enhancement, $|f|$. (d) Experimental and (e) numerically calculated phase spectra, $\varphi_3$, (solid lines). Lower panel shows the baseline-corrected spectra, $\Delta\varphi_3 = \varphi_3 - \varphi_3^{\text{bkg}}$ (curves offset for clarity), where $\varphi_3^{\text{bkg}}$ was obtained by assuming a non-absorbing object. The phase of the object polarizability, $\text{Arg}\{\alpha_O\}$, is shown for reference (black line). For panel (e) only: Dashed lines show the phase of the AOA term, $\varphi_{\text{AOA}}$, of the scattering model (Eq.19). (f) Parametric representation of the spectral signature of the molecular vibration (Si-CH$_3$) of PDMS, $\varphi_3^{\text{pp}}$, as obtained from panels (d,e), plotted against the calculated field enhancement, $f$. Error bars: see Methods.

Figure 3(a) shows the near-field amplitude spectra obtained with single gold nanorods, $\tilde{s}_3 = s_3 e^{i\varphi_3}$. By changing the nanorod length, we could control the antenna resonance with respect to the 1,258 cm$^{-1}$ (Si-CH$_3$) vibration of the PDMS coating on the near-field probe and bring about different tuning configurations: molecular vibration below resonance (red line), near-resonance (blue) and above resonance (green). The near-field amplitude spectra (solid lines in Fig. 3(a)) show the typical fundamental resonance of the nanorods that follows a linear scaling with the nanorod length according to $L_{\text{res}} = \lambda_{\text{eff}}/2$, where $\lambda_{\text{eff}}$ is the effective wavelength for the metal nanorod[26]. The spectral signature of the vibrational resonances of PDMS is clearly seen at the marked frequencies (arrows in Fig. 3(a)). Interestingly, the observed spectral signature is very different compared to the Fano-like shape of typical SEIRA spectra. Specifically, a dispersive line shape is observed when the antenna is tuned to a molecular vibration (blue line) and a step-like shape in the below-resonance case (green line). We isolated the spectral signature of the molecular vibrations by (i) calculating the scattering spectra assuming a non-absorbing object, $s_3^{\text{bkg}}$ (shaded area in Fig. 3(a), using the scattering model below), and (ii) subtracting the calculated spectra from the experimental scattering spectra, $\Delta s_3 = s_3 - s_3^{\text{bkg}}$ (Fig. 3(a), bottom panel, see Methods). We found that the spectral signature of the molecular vibration is clear dispersive for all considered rod lengths and resembles the magnitude of the polarizability, $|\alpha_O|$ (black line in Fig. 3(b), bottom). This observation provides clear evidence that the interaction between antenna and molecular vibration is a scattering process, and that field-enhanced molecular scattering can be measured and is a significant quantity.

To corroborate the experimental data, we performed numerical calculations of the presented near-field experiment based on finite-difference time-domain (FDTD) calculations[27,28]. These calculations describe (i) the near-field coupling between antenna and object, (ii) the optical apparatus for collection and refocusing of the scattered light from the sample, and (iii) signal demodulation and interferometric detection, thus providing a comprehensive description of the experimentally recorded data (see Methods and Extended Data Figure 4). The numerical calculations reproduced qualitatively and in very good agreement the experimental data, and particularly, the shape of the spectral signature of the molecular vibrations was correctly reproduced (cf. solid lines in Figs. 3(b) and 3(a)). Further, we compared the field-enhanced molecular scattering, $|E_{\text{AOA}}|$, from the scattering model (Eq. 18) (dashed lines) with the calculated spectra (solid lines in Fig. 3(b)). Remarkably, the scattering model is capable of reproducing the main spectral features including the spectral signature of the molecular vibration for all three antennas.

To relate the spectral signature of the Si-CH$_3$ molecular vibration with the local field enhancement of the antenna, we extracted the peak-to-peak value of spectral signature of the molecular vibration, $\Delta s_3^{\text{pp}}$, as a function of nanorod length, $L$, and compared it against the calculated field enhancement, $|f|$. Figure 3(c) shows a parametric representation of the $\log \Delta s_3^{\text{pp}}(L)$ and $\log|f(L)|$. With the experimental near-field spectra, we found a nearly linear relationship, revealing that the spectral signature, $\Delta s_3^{\text{pp}}$, and field enhancement, $|f|$, are related by a power law (red points in Fig. 3(c)). Performing a linear least-square fitting, we obtained a nearly quadratic scaling of the spectral signature with $\Delta s_3^{pp} \propto |f|^{1.8 \pm 0.4}$, which is close to the expected power-of-two law of field-enhanced molecular scattering, $|\mathbf{E}_{AOA}| \propto |f|^2$, as described in Eq. (18). The corresponding intensity of the spectral signature thus scales with $\left(\Delta s_3^{pp}\right)^2 \propto |f|^{3.60}$, confirming that the spectral signature of field-enhanced molecular scattering scales with the fourth power of the local field enhancement. We obtained a similar scaling law of $\Delta s_3^{pp} \propto |f|^{1.94 \pm 0.09}$ with the data of the numerical model (black points in Fig. 3(c)), which further supports our experimental data. The discrepancy between the experimental and numerical data can be attributed to experimental noise, which rendered measurements of antennas that were far off resonance with the molecular vibration (small $|f|$) less reliable.

In addition, the near-field phase spectra, $\varphi_3$, in Figs. 3(d,e) show the typical response of an field-enhanced molecular scattering process with a transition from 0° to 360° as the fundamental resonance of the antenna is crossed. More precisely, the typically response of an antenna (phase transition from 0° to 180° across the antenna resonance) is doubled owing to the double-scattering nature of the AOA term.[17] The spectral signature of the molecular vibration appears in form of a positive peak that sits on top of the slowly varying phase response of the antenna. By extracting the baseline-corrected phase spectra, $\Delta \varphi_3 = \varphi_3 - \varphi_3^{bkg}$, we found that the spectral signature of the molecular vibration had the shape of the argument of the polarizability, $\text{Arg}\{\alpha_O\}$, (bottom panel in Figs. 3(d,e)). We further found that the peak height, $\Delta \varphi_3^{pp}$, of the Si-CH$_3$ molecular vibration was approximately independent of the local field enhancement, $f$, at a value of about 21° (data points in Fig. 3(f)), which agrees well with the corresponding peak height in the argument of the polarizability, $\text{Arg}\{\alpha_O\}$ (dashed line in Fig. 3(f)), and is consistent with phase of the field-enhanced molecular scattering as described in Eq. (19).

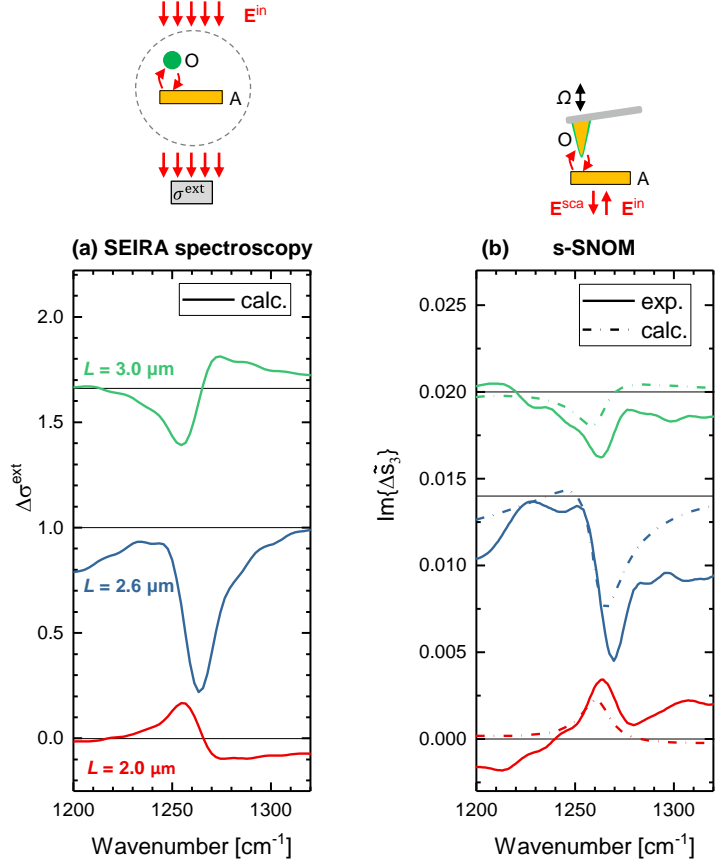

**Figure 4: Field-enhanced molecular scattering as measured by s-SNOM yields the vibrational line shapes in the extinction cross section.** (a) Calculated spectral signature of the molecular vibrations in the extinction cross section, $\Delta\sigma^{\text{ext}} = \sigma^{\text{ext}} - \sigma_{\text{bkg}}^{\text{ext}}$, of the model system of a nanorod of different lengths $L$ coupling to a PDMS nanoparticle. (a) Imaginary part, $\text{Im}\{\Delta\tilde{s}_3\}$, of the baseline-corrected near-field spectra, $\Delta\tilde{s} = \tilde{s}_3 - \tilde{s}_3^{\text{bkg}}$ (solid lines: using the experimental s-SNOM data from Fig. 3(a,d), dashed lines: using the calculated s-SNOM data from Fig. 3(b,e)). The superscript bkg indicates that a non-absorbing object is considered.

In Fig. 4 we establish a connection between the near-field experiment shown in Fig. 2 with the vibrational line shapes observed in the extinction cross section. To this end, we plot the baseline-corrected extinction cross section, $\Delta\sigma^{\text{ext}} = \sigma^{\text{ext}} - \sigma_{\text{bkg}}^{\text{ext}}$, in Fig. 4(a), where $\sigma^{\text{ext}}$ and $\sigma_{\text{bkg}}^{\text{ext}}$ is the extinction cross section for an absorbing and a non-absorbing object, respectively. This baseline correction effectively removes the direct antenna scattering from $\sigma^{\text{ext}}$ ($\mathbf{E}_A$ in Eq. (9)) and thus isolates the spectral signature of the molecular vibration. For comparison, we plot the imaginary part of the baseline-corrected near-field spectra, $\Delta\tilde{s}_3 = \tilde{s}_3 - \tilde{s}_3^{\text{bkg}}$, in Fig. 4(b), where $\tilde{s}_3$ is the measured near-field spectrum (solid lines in Fig. 3(a,d)) and $\tilde{s}_3^{\text{bkg}}$ is the calculated near-field spectrum assuming a near-field probe that is covered by non-absorbing dielectric thin film (shaded area in Fig. 3(a)). This baseline correction is motivated as follows: While the direct antenna scattering $\mathbf{E}_A$ is responsible for the baseline effect in the extinction cross section, $\mathbf{E}_A$ is suppressed in the near-field experiment by the demodulation techniques. Instead, it is the field-enhanced scattering from the metal core of the near-field probe that produces a strong baseline effect in the s-SNOM experiment, yielding the antenna resonance curve (the shaded area in Fig 3(a)). We note that the polarizability of the PDMS-covered near-field probe may be separated into the sum of background (describing essentially the metal core) and vibrational components (describing the molecular vibration), which can be shown by developing Eq. (36) in the limit of a weak molecular oscillator. Further noting that $\tilde{s}_3 \propto \mathbf{E}_{\text{AOA}} \propto \alpha_O$ (Eqs. (6), (17)), the field-enhanced molecular scattering can be isolated by taking the complex-valued difference, $\Delta \tilde{s}_3 = \tilde{s}_3 - \tilde{s}_3^{\text{bkg}}$. The imaginary part of $\Delta \tilde{s}_3$ can then be calculated straightforwardly, which is motivated by Eq. (9), and possible because the antenna-scattered light is resolved in amplitude and phase and normalized to the incident field $\mathbf{E}_{\text{in}}$ (Methods). The result is a good agreement of the s-SNOM data with the vibrational line shape in the extinction cross section $\Delta\sigma^{\text{ext}}$ (cf. Fig. 4(a,b)), thus providing experimental evidence that the interference between field-enhanced molecular scattering $\mathbf{E}_{\text{AOA}}$ and the incident field can fully explain the vibrational signature in SEIRA spectra.

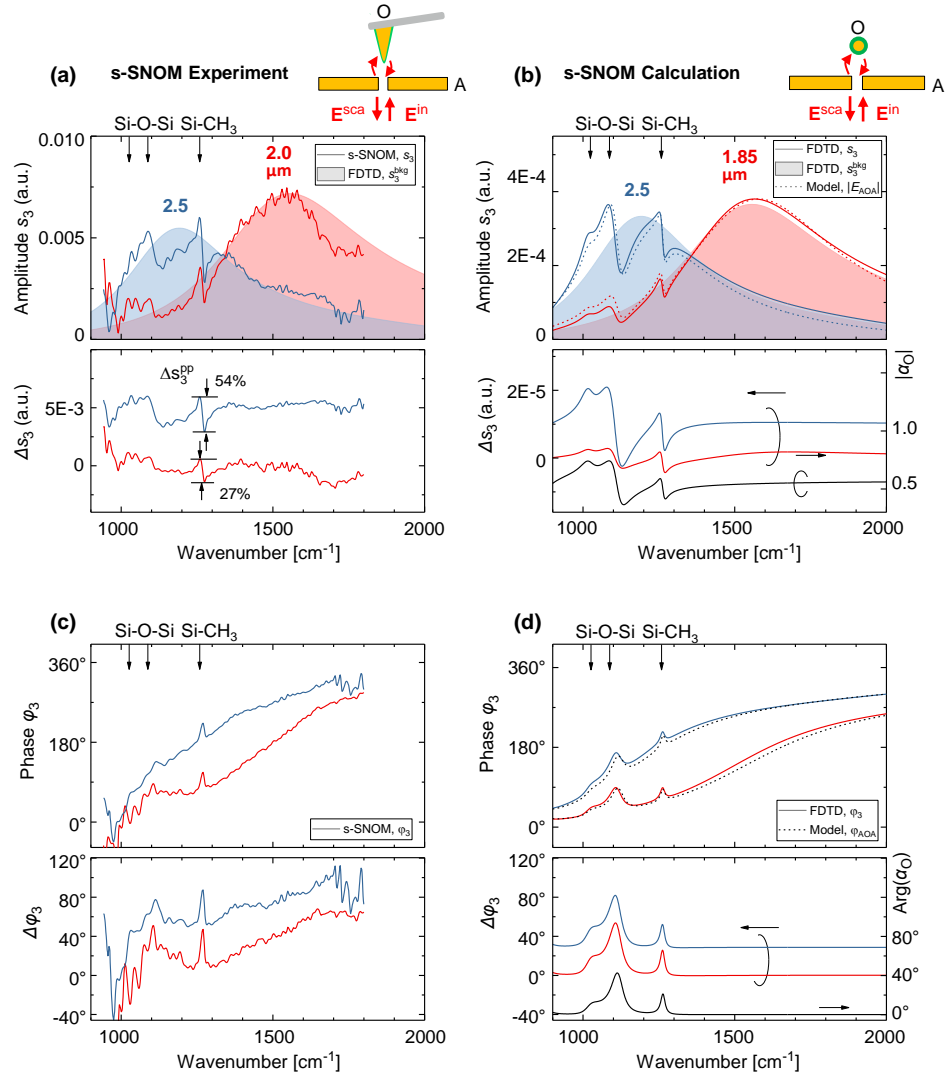

**Figure 5: Near-field spectroscopic measurement of field-enhanced molecular scattering with a gap antenna.** (a) Experimental and (b) numerically calculated amplitude spectra, $s_3$, of a gap antenna of different lengths, $L$ (solid lines). Shaded areas show the calculated near-field spectra, $s_3^{bkg}$, assuming a non-absorbing object. Bottom panel shows the isolated spectral signature of the molecular vibrations, $\Delta s_3 = s_3 - s_3^{bkg}$, (curves offset for clarity). The magnitude of the object polarizability, $|\alpha_O|$, is shown for reference (black line). (c) Experimental and (d) numerically calculated phase spectra, $\varphi_3$, (solid lines). Lower panel shows the baseline-corrected spectra, $\Delta\varphi_3 = \varphi_3 - \varphi_3^{bkg}$ (curves offset for clarity), where $\varphi_3^{bkg}$ was obtained by assuming a non-absorbing object. The phase of the object polarizability, $\text{Arg}\{\alpha_O\}$, is shown for reference (black line). Panels (b),(d): Dashed lines show the magnitude and phase of the AOA term, $|E_{AOA}|$ and $\text{Arg}\{\alpha_O\}$, of the scattering model (Eq. 18).

## Discussions & Conclusions

We have demonstrated theoretically and numerically that vibrational line shapes in the extinction cross section can be explained by the interference of field-enhanced molecular scattering $\mathbf{E}_{\text{AOA}}$ with the incident field $\mathbf{E}^{\text{in}}$ when molecular vibrations couple to a single antenna. Our experimental study shows that field-enhanced molecular scattering $\mathbf{E}_{\text{AOA}}$ can be measured, scales in intensity with the fourth power of the local field enhancement, $|f|^4$, and fully explains the vibrational signature in SEIRA spectra.

A key feature of our results is the large spectral signature of the molecular vibrations in the experimental amplitude spectra, $|E_n|$. With the nanorods shown in Fig. 3, the spectral signature can reach up to 41% of antenna resonance maximum (Fig. 3(a), bottom). With a gap antenna – affording stronger coupling between molecule and antenna – an even larger spectral signature can be observed of up to 54% (Fig. 5(a), bottom), which is a result of the demodulation techniques employed in the presented s-SNOM experiment. Thus, the combination of scattered light detection and demodulation techniques could be exploited to build highly sensitive SEIRA sensors.

Finally, we have described the development of a nano-FTIR transflection-mode setup for the spectroscopic mapping of resonant antenna structures in the near field, as well as a numerical model that can accurately predict the experimental near-field spectra, which is a useful tool for studying coupling phenomena on the nanoscale between individual objects. Particularly, near-field microscopy using functionalized tips is demonstrated as a useful tool to map SEIRA enhancement in nanophotonic structures.

**Funding.**

This work was financially supported by: Grant CEX2020-001038-M and grant PID2020-115221GA-C44 (SNOMCELL) funded by MICIU/AEI /10.13039/501100011033. Grant PID2021-123949GB-I00 (NANOSPEC) funded by MICIU/AEI /10.13039/501100011033 and by ERDF/EU.

**Acknowledgements.**

The authors thank Irene Dolado Lopez for providing the antenna sample and Javier Aizpurua, Ilia Rasskazov and Scott P. Carney for fruitful discussions.

**Author contributions.**

D. V. performed the measurements. M. S. performed the numerical calculations. D. V., C. M.-E. and M. S. performed data analysis. C. M.-E., R. H. and M. S. developed the theory. R. H. and M. S. wrote the manuscript with input from all authors. All authors contributed to scientific discussions.

**Competing interests.**

R.H. is co-founder of Neaspec GmbH, a company producing scattering-type scanning near-field optical microscope systems, such as the one used in this study. The remaining authors declare no competing interests.

**Correspondence** and requests for materials should be addressed to Martin Schnell.

**Data availability.**

Data underlying the results presented in this paper are not publicly available at this time but may be obtained from the authors upon reasonable request.

## METHODS

### Elastic Light Scattering Process

We describe the scattered field of the antenna-object system $\mathbf{E}^{sca}$ as a series of multiple scattering events occurring between the antenna and the object in the limit of a weakly scattering object (i.e. small object polarizability, $\alpha_O$) [18,19].

$$\mathbf{E}^{sca}(\mathbf{r}) = \mathbf{E}_A(\mathbf{r}) + \mathbf{E}_O(\mathbf{r}) + \mathbf{E}_{OA}(\mathbf{r}) + \mathbf{E}_{AO}(\mathbf{r}) + \mathbf{E}_{AOA}(\mathbf{r}) + \cdots. \qquad (20)$$

The following expressions may be obtained for the individual terms by considering the tensorial nature of the diagonal antenna polarizability, $\overleftrightarrow{\alpha}_A$, the object polarizability, $\overleftrightarrow{\alpha}_O$, and the field enhancement, $\overleftrightarrow{f}$ (tensor of components, $f_{ij}$ with $i,j = x,y,z$) [17,19]:

$$\mathbf{E}_A(\mathbf{r}) = k^2 \overleftrightarrow{\mathbf{G}}(\mathbf{r},\mathbf{r}_A) \overleftrightarrow{\alpha}_A \mathbf{E}^{in}(\mathbf{r}_A), \qquad (21)$$

$$\mathbf{E}_O(\mathbf{r}) = k^2 \overleftrightarrow{\mathbf{G}}(\mathbf{r},\mathbf{r}_O) \overleftrightarrow{\alpha}_O \mathbf{E}^{in}(\mathbf{r}_O), \qquad (22)$$

$$\mathbf{E}_{OA}(\mathbf{r}) = k^2 \overleftrightarrow{\mathbf{G}}(\mathbf{r},\mathbf{r}_O) \overleftrightarrow{\alpha}_O \overleftrightarrow{f} \mathbf{E}^{in}(\mathbf{r}_A), \qquad (23)$$

$$\mathbf{E}_{AO}(\mathbf{r}) = k^2 \overleftrightarrow{\mathbf{G}}(\mathbf{r},\mathbf{r}_A) \overleftrightarrow{f}^T \overleftrightarrow{\alpha}_O \mathbf{E}^{in}(\mathbf{r}_O), \qquad (24)$$

$$\mathbf{E}_{AOA}(\mathbf{r}) = k^2 \overleftrightarrow{\mathbf{G}}(\mathbf{r},\mathbf{r}_A) \overleftrightarrow{f}^T \overleftrightarrow{\alpha}_O \overleftrightarrow{f} \mathbf{E}^{in}(\mathbf{r}_A), \qquad (25)$$

where $\mathbf{r}_A$ and $\mathbf{r}_O$ are the position of the dipole that represents the antenna and object, respectively, and $\mathbf{r}$ is the point where the electric field evaluated, $\overleftrightarrow{\mathbf{G}}$ is the Green's function in free-space. Note that the field enhancement tensor is defined as $\mathbf{f} = k^2 \mathbf{G}(\mathbf{r}_O, \mathbf{r}_A)\alpha_A$. Approximate expressions may be obtained by considering the configuration of the numerical calculations. First, the incident field is configured to be x-polarized, $\mathbf{E}^{in}(\mathbf{r}_A) = \mathbf{E}_0^{in} \exp(i\mathbf{k}\cdot\mathbf{r}_A) = E_x^{in}(\mathbf{r}_A)\hat{\mathbf{x}}$, and thus parallel to the long axis of the antenna considered in Fig. 1. Second, the electric field at the end and on the long axis of the rod antenna is mainly x-polarized (dominance of the one component of the field enhancement factor over the others, $|f_{xx}| \gg |f_{yx}|, |f_{zx}|$). Third, the polarizability of the spherical object is assumed to be isotropic ($\alpha_{O,xx} = \alpha_{O,yy} = \alpha_{O,zz}$), that is, we consider that the object consists of a large number of randomly oriented molecules. Above terms (Eqs. (21)-(25)) may thus be approximated to

$$\mathbf{E}_A(\mathbf{r}) \approx k^2 \mathbf{G}_x(\mathbf{r},\mathbf{r}_A) \alpha_A E_x^{in}(\mathbf{r}_A), \qquad (26)$$

$$\mathbf{E}_O(\mathbf{r}) \approx k^2 \mathbf{G}_x(\mathbf{r},\mathbf{r}_O)\alpha_O E_x^{in}(\mathbf{r}_O)\,, \tag{27}$$

$$\mathbf{E}_{OA}(\mathbf{r}) \approx k^2 \mathbf{G}_x(\mathbf{r},\mathbf{r}_O)\alpha_O f E_x^{in}(\mathbf{r}_A)\,, \tag{28}$$

$$\mathbf{E}_{AO}(\mathbf{r}) \approx k^2 \mathbf{G}_x(\mathbf{r},\mathbf{r}_A)f\alpha_O E_x^{in}(\mathbf{r}_O)\,, \tag{29}$$

$$\mathbf{E}_{AOA}(\mathbf{r}) \approx k^2 \mathbf{G}_x(\mathbf{r},\mathbf{r}_A)f\alpha_O f E_x^{in}(\mathbf{r}_A)\,, \tag{30}$$

where $\mathbf{G}_x$ is x-component of the Green's tensor function, $\mathbf{G}_x = \overleftrightarrow{\mathbf{G}} \cdot \hat{\mathbf{x}}$, and indices $xx$ were omitted for clarity. Considering the much smaller polarizability of the object compared to the antenna, $\alpha_A \gg \alpha_O$, the interaction series in Eq. (20) may be truncated to first order in O. Further assuming large field enhancement, $f \gg 1$, it is sufficient to only retain terms $\mathbf{E}_A(\mathbf{r})$ and $\mathbf{E}_{AOA}(\mathbf{r})$ in $\mathbf{E}^{sca}(\mathbf{r})$, where $\mathbf{E}_{AOA}(\mathbf{r})$ describes the field scattered from the object via the antenna after being illuminated by the antenna. Terms $\mathbf{E}_{OA}(\mathbf{r})$ and $\mathbf{E}_{AO}(\mathbf{r})$ – describing a single scattering event between antenna and object – only scale linearly in $f$ and are thus much smaller than the term $\mathbf{E}_{AOA}(\mathbf{r})$ and will be neglected. The scattered field may thus be reduced to

$$\mathbf{E}^{sca}(\mathbf{r}) \approx \mathbf{E}_A(\mathbf{r}) + \mathbf{E}_{AOA}(\mathbf{r}) \tag{31}$$

The scattered field intensity may be similarly approximated. With Eq. (20), $|\mathbf{E}^{sca}|^2$ may be developed as follows (argument $\mathbf{r}$ omitted for brevity):

$$\frac{|\mathbf{E}^{sca}|^2}{|\alpha_O|^2} \approx \underbrace{\mathbf{E}_A^* \cdot \mathbf{E}_A}_{|\alpha_A|^2 \sim 6\cdot 10^{19}} + \underbrace{\mathbf{E}_A^* \cdot \mathbf{E}_{AOA}}_{|\alpha_A f^2 \alpha_O| \sim 2\cdot 10^{13}} + \underbrace{\mathbf{E}_A^* \cdot \mathbf{E}_{AO}}_{|\alpha_A f \alpha_O| \sim 4\cdot 10^{11}} + \underbrace{\mathbf{E}_A^* \cdot \mathbf{E}_{OA}}_{|\alpha_A f \alpha_O| \sim 4\cdot 10^{11}} + \underbrace{\mathbf{E}_A^* \cdot \mathbf{E}_O}_{|\alpha_A \alpha_O| \sim 8\cdot 10^{9}} + c.c., \tag{32}$$

where the series was truncated to first order in O and terms are listed in descending order of importance and $c.c.$ means complex conjugate. By way of example, an estimate for the magnitude is stated below each term in units of $|\alpha_O|^2$ (pure molecular scattering), assuming the values from the numerical calculation in main text (1 nm radius spherical nanoparticle), $2|\alpha_A|/|\alpha_O| \sim 1.6 \cdot 10^{10}$ and $f \sim 50$. It is apparent that above series may be reduced by retaining only the first and second term, while incurring an error of only about 4% by omitting terms 3 and 4:

$$|\mathbf{E}^{sca}(\mathbf{r})|^2 \approx \mathbf{E}_A^*(\mathbf{r}) \cdot \mathbf{E}_A(\mathbf{r}) + 2\mathrm{Re}\{\mathbf{E}_A^*(\mathbf{r}) \cdot \mathbf{E}_{AOA}(\mathbf{r})\}\,, \tag{33}$$

Note that so far, we have assumed that the object (O) is positioned on the long axis of the nanorod antenna, as it is the case in Fig. 1. If the object (O) is located on top of the nanorod antenna (as it is the case in Fig. 3), then we can make use of the fact that the z-component of the field enhancement is dominant near the nanorod ends, $|f_{zx}| \gg |f_{xx}|, |f_{yx}|$. We can then approximate $\mathbf{E}_{\text{AOA}}$ by

$$\mathbf{E}_{\text{AOA}}(\mathbf{r}) \approx k^2 \mathbf{G}_x(\mathbf{r}, \mathbf{r}_A) f_{zz} \alpha_{O,zz} f_{zz} E_x^{\text{in}}(\mathbf{r}_A) , \qquad (34)$$

By substituting $\alpha_O = \alpha_{O,zz}$ (spherical object) and $f = f_{zz}$, we arrive at the same expression for the scattered field, $\mathbf{E}^{\text{sca}}$.

**Numerical Calculation of SEIRA**

We provide details on the numerical study presented in Fig. 1, Extended Data Table 1 and Extended Data Figs. 1-3. We considered a cylindrical gold antenna of 50 nm radius and with round end caps. The antenna was placed in vacuum, that is, no substrate was assumed. A spherical nanoparticle (NP) with 10 nm radius was placed on the antenna's long axis with a separation of 20 nm between the antenna end cap and the nanoparticle surface. We used the Lorentzian model to describe the vibrational resonance of the NP:

$$\varepsilon_O(\omega) = \varepsilon_{\text{bkg}} + \varepsilon_{\text{vib}} = \varepsilon_{\text{bkg}} + \frac{\varepsilon_{\text{Lorentz}} \cdot \omega_0^2}{\omega_0^2 - \omega^2 - 2i\gamma\omega}, \qquad (35)$$

with background permittivity $\varepsilon_{\text{bkg}} = 1.55$, and resonance $\omega_0 = 1{,}258$ cm$^{-1}$ and linewidth $\gamma = 10$ cm$^{-1}$. A Lorentz oscillator strength $\varepsilon_{\text{Lorentz}} = 0.015$ was assumed to mimic the Si-CH$_3$ vibrational resonance of PDMS that is probed in the experiments. A larger Lorentz oscillator strength $\varepsilon_{\text{Lorentz}} = 0.1$ and a NP radius of 30 nm was assumed in Fig. 1 to show more clearly the spectral signature of the molecular vibration. We assumed tabulated values for the permittivity of Au[29] for the antenna. We used a commercial software package based on the Finite-difference time-domain (FDTD) method to calculate the relevant quantities of SEIRA (Ansys Lumerical FDTD, Ansys, Inc). For each configuration of the antenna-NP system, we performed a total of four calculations. In all calculations, plane-wave illumination of the antenna with the polarization along the antenna axis was assumed, and the scattering cross section was calculated by considering the outward flowing net power. A mesh of 5 nm was assumed around the antenna and the mesh was refined at the NP to 1 nm (NP radius $\geq$ 10 nm) and 0.5 nm (NP radius < 10 nm). First, we calculated the extinction and scattering cross section, $\sigma^{\text{ext}}$ and $\sigma^{\text{sca}}$, of the antenna-NP system by considering an absorbing NP (Eq. (35)). Second, we calculated the extinction and scattering cross section, $\sigma_{\text{bkg}}^{\text{ext}}$ and $\sigma_{\text{bkg}}^{\text{sca}}$, of the antenna-NP system by considering a non-absorbing NP ($\varepsilon_{\text{Lorentz}} = 0$ in Eq. (35)). From these two calculations we obtained the spectral signature of the molecular vibration in extinction, $\Delta\sigma^{\text{ext}} = \sigma^{\text{ext}} - \sigma_{\text{bkg}}^{\text{ext}}$, and scattering cross section, $\Delta\sigma^{\text{sca}} = \sigma^{\text{sca}} - \sigma_{\text{bkg}}^{\text{sca}}$. Third, we calculated the 3D near-field distribution of the antenna only (i.e. the NP was removed). From this recorded near-field distribution we obtained the dipole moment of the antenna, $p_A$, following the approach from ref. [10]. Fourth, we determined the incident field at the center of the antenna, $E_x^{\text{in}}$. We then calculated the antenna polarizability, $\alpha_A = p_A/E_x^{\text{in}}$. From the recorded near-field distribution (3$^{\text{rd}}$ simulation), we further obtained the x-component of the near fields produced by the antenna at the center of the NP, $E_{\text{NF},x}$, from which we calculated the corresponding field enhancement factor, $f = E_x/E_x^{\text{in}}$ (the *y* and *z*-components can be neglected, see section above). Note that for NP radius of 10 nm or lower, it is sufficient to evaluate the field enhancement at the center of the NP. Only in case of NP radius of 30 nm, where the inhomogeneity of the antenna near fields become significant, the average field enhancement factor $f_{\text{avg}}^2 = \frac{1}{V_{\text{np}}} \int_{V_{\text{np}}} f^2 \, dV$ was calculated.[15] Finally, we evaluated the object polarizability analytically according to $\alpha_O = 4\pi a^3 \frac{\varepsilon_O - 1}{\varepsilon_O + 2}$, where $a$ is the radius and $\varepsilon_O$ is the permittivity of the object. We then straightforwardly calculated the spectral signature of the molecular vibration from the scattering model, $\sigma_{\text{vib}}^{\text{ext}}$ and $\sigma_{\text{vib}}^{\text{sca}}$, by inserting above quantities in Eq. (13).

We note that small antenna geometries such as the rod antenna considered here are accurately described by their dipole moment (to less than 1% error in scattering). This approximation may prove to be insufficient for a quantitative prediction of SEIRA in case of larger antenna structures, requiring an adaptation of the scattering model. A possible approach is outlined in ref. 15. Nevertheless, it was shown that the picture of the point polarizability may still provide a qualitative description of the scattering of large antenna geometries (such as in near-field microscopy[18,30]), possibly extending the applicability of the scattering model to qualitatively describe SEIRA in this regard. We further note that the presented scattering model applies for both thick (scattering dominated) and thin (absorption dominated) rod antennas because knowledge of the forward scattered field alone is sufficient to evaluate the extinction cross section with the optical theorem, and such variation of the antenna polarizability $\alpha_A$ will be explored in the future.

**Numerical Calculation of the Near-Field Experiment**

We provide details on the numerical calculations presented in Figs. 3,5. We considered rectangularly shaped Au nanorods (the antenna) of dimension *L* x 250 nm x 60 nm, where length *L* is variable as stated in Fig. 3(b), while width and height were common for all antennas and were determined from the topography data of the nanorods used in the experiment. The nanorod length *L* was adjusted to fit the experimental spectral in Fig. 3(a). The nanorod is supported on a CaF$_2$ substrate. In the experiment, the near-field probe was a pyramidically-shaped metallic structure with a nominal height in the range of 10 to 15 µm. The near-field probe was covered by a thin PDMS layer, a well known contamination of commercial AFM probes[23,24]. We modeled the actual near-field probe as a spherical Au nanoparticle (the object), where the radius was assumed to be equal to the radius of the tip apex of the near-field probe, that is 50 nm. We additionally assumed a 10 nm thin PDMS coating of the nanoparticle to mimic the PDMS contamination of the experimental near-field probe. The thickness of the PDMS coating was chosen, and the tapping amplitude was adjusted to 80 nm, to match the magnitude of the spectral signature observed in the experiment. We used tabulated values for the permittivity of Au[29], PDMS[31] and CaF$_2$[29]. In Fig. 5, we followed the same approach for the gap antenna by considering two rectangular shaped Au nanorods (the antenna) of dimension $L$ x 250 nm x 60 nm and separated by a 100 nm gap with the near-field probe being placed in the gap.

The scattered signal of the antenna-object system was calculated as follows (solid lines in Fig. 3(b,e)). We based the scattering model on the numerical algorithm described by Çapoğlu et al. [27] for implementing a virtual imaging system on a computer and adapted it to describe the specifics of our near-field experiment in Fig. 3. In detail, nanorod (antenna) and near-field probe (object) were illuminated with a focused Gaussian beam as injected with source S inside the FDTD box (Extended Data Fig. 4(a)). The optical response of this system was obtained by rigorously solving Maxwell's equation using the FDTD method. The near-field of the antenna-object system, $E_{\mathrm{NF}}(x,y)$, was recorded with monitor M inside the FDTD box and then subsequently propagated to the far field using a near-field-to-far-field transform to obtain the scattered field from the antenna-object system, $E_{\mathrm{FF}}(s_x, s_y)$. Scattering outside the objective NA was truncated and the remaining field was refocused onto the image (detector) plane, yielding electric field $E_{\mathrm{IM}}(x', y')$. For simplicity, a magnification of $M = 1$ was assumed.

Extended Data Fig. 4(b) illustrates the data processing flow. To simulate the demodulation process in s-SNOM, a series of simulations was run where the object-antenna distance, $d$, was varied linearly in steps of 5 nm, yielding a total of 35 electric field maps $E_{\mathrm{IM}}(x', y')$, one for each position of the tip. Note that a fixed mesh refinement covering the nanorod (antenna) and the small sphere (object) of 5 nm was assumed for all positions of the sphere (to avoid any residual effects caused by different mesh configurations in the demodulation process) and further a minimum distance of 2 mesh cells between nanorod and sphere was assumed (to avoid erroneous results when sphere and nanorod are in contact). These field maps, $E_{\mathrm{IM}}$, were then processed following the procedure from ref. [28]. First, interferometric detection was implemented by calculating the field of the reference beam at the image plane, $E_{\mathrm{IM}}^{\mathrm{ref}}$, as obtained by reflecting the focused beam in the simulation at a metal interface. Phase-shifting interferometry was applied to resolve the collected scattered field, $E^{\mathrm{sca}}$, in amplitude and phase. Extended Data Fig. 4(c) exemplarily shows the collected scattered field, $E^{\mathrm{sca}}$, for a few selected tip positions. The antenna resonance is clearly visible, but appears as a negative peak owing to interference with the reflected light off the substrate surface. For clarity, we remove the reflection off the substrate surface and the static antenna scattering, $E_A$, by running a simulation where the object is removed. Extended Data Fig. 4(d) shows the resulting differential spectra, $\Delta E^{\mathrm{sca}} = E^{\mathrm{sca}} - E_A$, revealing the spectral signature of PDMS. Extended Data Fig. 4(e) shows the normalized differential spectra, $\Delta E^{\mathrm{sca}}$, showing that the spectral shape of $\Delta E^{\mathrm{sca}}$ is nearly constant as the object approaches the antenna, confirming that modulation of the object-antenna distance, $d$, only enters as a modulation of the field enhancement, $f = f(d)$, but does not affect the spectral shape of the object-antenna coupling, $E_{\mathrm{AOA}}$, as expected[17]. We then constructed the sinusoidal motion of the s-SNOM probe in the experiment, $d = d_0 + \Delta d \cos \Omega t$ (where $\Delta d$ is the modulation amplitude and $d_0$ the average tip height), by interpolating the scattered field, $E^{\mathrm{sca}}(d)$, at a total of 100 positions in $d$. A tapping amplitude of 80 nm was assumed. The so obtained complex-valued scattered field, $E^{\mathrm{sca}}(d)$, was demodulated at the 3rd harmonic, $3\Omega$, to yield the calculated near-field, $E_3^{\text{raw}}$. Signal normalization was performed as it was done in the experiment by normalizing the demodulated spectra, $E_3^{\text{raw}}$, to the reference field, $E^{\text{ref}}$, yielding the normalized demodulated scattering signal $\tilde{s}_3 = s_3 e^{i\varphi_3} = E_3^{\text{raw}}/E^{\text{ref}}$. The corresponding normalized amplitude and phase spectra, $s_3(\omega)$ and $\varphi_3(\omega)$, are shown as solid curves in Fig. 3(b).

For comparison with the simulation assuming an absorbing object (PDMS), we repeated above calculation with a non-absorbing object, where we assumed a background permittivity of $\varepsilon_{\text{bkg}} = 1.8$, yielding the demodulated scattered field, $\tilde{s}_3^{\text{bkg}} = s_3^{\text{bkg}} e^{i\varphi_3^{\text{bkg}}}$ (shaded curves in Fig. 3(b)). With the help of these spectra, isolation of the spectral signature of the molecular vibration of the numerical data could be done straightforwardly by taking the following difference: $\Delta s_3 = s_3 - s_3^{\text{bkg}}$ (isolating the spectral signature in amplitude) and $\Delta \varphi_3 = \varphi_3 - \varphi_3^{\text{bkg}}$ (isolating the spectral signature in phase). To perform this isolation with the experimental data, we used the calculated spectra, $s_3^{\text{bkg}} e^{i\varphi_3^{\text{bkg}}}$, as a substitute for experimental spectra of a non-absorbing object (which would be difficult to do experimentally because it would require preparation of s-SNOM tips with a non-absorbing dielectric film of equal thickness compared to the PDMS contamination). To this end, we scaled the calculated spectra as follows. The calculated spectra assuming an absorbing object, $s_3 e^{i\varphi_3}$ (red line in Extended Data Fig. 5), were scaled to the maximum of the experimental spectra (black line). Using the same scaling factor, we then plotted the calculated spectra assuming a non-absorbing object, $s_3^{\text{bkg}} e^{i\varphi_3^{\text{bkg}}}$ (blue line in Extended Data Fig. 5 and shaded area in Fig. 3(a)). After this scaling, we could isolate the spectral signature of the molecular vibrations in the experimental data by straightforwardly calculate the differences, $\Delta s_3 = s_3 - s_3^{\text{bkg}}$ and $\Delta \varphi_3 = \varphi_3 - \varphi_3^{\text{bkg}}$.

To evaluate the field-enhanced molecular scattering, $\mathbf{E}_{\text{AOA}}$, from the scattering model (Eqs. 18, 19) (plotted as dashed lines in Fig. 3(b,e)), we calculated the field enhancement, $f \approx f_z$, for the unloaded rod antenna at the lowest position of the particle, d, where we only consider the z-component as it is the dominant component over the x- and y-components above the metal nanorod. Further, we evaluated the polarizability of the Au nanoparticle, $\alpha_O$, with a PDMS coating by using the analytical model for a core-shell nanoparticle in the quasistatic limit[22]:

$$\alpha_O = 4\pi R_2^3 \frac{R_2^3(\varepsilon_2 - 1)(\varepsilon_1 + 2\varepsilon_2) + R_1^3(\varepsilon_1 - \varepsilon_2)(1 + 2\varepsilon_2)}{R_2^3(\varepsilon_2 + 2)(\varepsilon_1 + 2\varepsilon_2) + 2R_1^3(\varepsilon_2 - 1)(\varepsilon_1 - \varepsilon_2)}, \qquad (36)$$

where we assumed a radius for the metallic core of $R_1 = 50$ nm and a 10 nm thick PDMS coating, i.e. $R_2 = 60$ nm, and permittivity for Au ($\varepsilon_1$) and PDMS ($\varepsilon_2$) as in the numerical calculation above. We note that both the metal core and molecular layer contribute to the scattered light, $E_{AOA}$, in that scattering by the metal core via the antenna after being illuminated by the antenna essentially probes the field enhancement of the antenna[17], which is the reason why the antenna resonance is seen so clearly in Fig. 3(b). Conversely, scattering via the molecular layer via the antenna after being illuminated by the antenna probes the molecular vibration. Both scattering contributions add up to yield the spectra shown in Fig. 3(b) and are modelled here by introducing an effective object polarizability, $\alpha_O$, as defined above in Eq. (36). From the good agreement with the experimental data in Fig. 3(a), we conclude that in our specific experiment the molecule-coated tip is well described by a core-shell nanoparticle.

**Sample Fabrication**

We fabricated Au rods on the CaF$_2$ substrate via high-resolution electron-beam lithography. Polymethyl methacrylate (PMMA) was spin coated onto the substrate at 4000 rpm as the electronsensitive polymer. The PMMA was subsequently covered by a 2 nm thick layer of gold for enabling lithography on the insulating substrate. After the electron-beam assisted writing of the rods, the gold was chemically etched (5 s immersion in KI/I2 solution) and the PMMA was developed in methyl isobutyl ketone: isopropanol 1:3. Finally 5 nm layer of Ti were deposited by electron beam evaporation followed by thermal evaporation of 50 nm of Au. The lift-off of the rods was done by immersing the sample in acetone overnight.

**Near-Field Experiment**

We modified a commercial s-SNOM system (NeaSCOPE, attocube systems AG) to allow for sample illumination and light collection from below the sample (transflection-mode s-SNOM). This modality was recently demonstrated for near-field mapping of IR-resonant antennas (in liquid)[25]. The transflection-mode geometry allows for efficient excitation of the antenna structures on the sample. At the same time, direct excitation of the s-SNOM probe is largely avoided (because the long axis is oriented in the direction of light propagation), allowing for the use of metallic near-field probes as strong scatterer of local near fields. Here, we introduce and describe spectroscopic transflection-mode measurements of single IR-resonant antennas based on nano-FTIR. In detail, we used standard Pt-Ir coated (metallic) AFM tips (NCPt arrow tip, Nanoworld) to map the rod antennas. A parabolic mirror (focal length 8 mm, NA 0.44) was used to focus the broadband infrared beam of the nano-FTIR laser at normal incidence through a CaF$_2$ window (sample holder) and the substrate. The polarization of the illuminating beam was chosen to be parallel to the nanorod axis for efficient excitation of the dipolar resonance. The antenna scattered light was collected from below the sample with the same parabolic mirror, analyzed with the interferometer of the nano-FTIR module and detected with an MCT detector (InfraRed Associates, Inc.). Specifically, the antenna scattered field $\mathbf{E}^{sca}$ was superposed with the external reference field of nano-FTIR interferometer $\mathbf{E}^{ref}$, yielding a detector signal proportional to $I_d \propto |\mathbf{E}^{sca}|^2 + |\mathbf{E}^{ref}|^2 + 2\text{Re}\{\mathbf{E}^{ref*} \cdot \mathbf{E}^{sca}\}$. With Eqs. (3),(4), the detector signal can be approximated as

$$I_d \propto |\mathbf{E}_A|^2 + 2\text{Re}\{\mathbf{E}_A^* \cdot \mathbf{E}_{AOA}\} + 2\text{Re}\{\mathbf{E}^{ref*} \cdot \mathbf{E}_A\} + 2\text{Re}\{\mathbf{E}^{ref*} \cdot \mathbf{E}_{AOA}\} + |\mathbf{E}^{ref}|^2. \quad (37)$$

To suppress the direct antenna scattering, $\mathbf{E}_A$, in Eq. (37), the near-field probe was vertically vibrated sinusoidally with amplitude $\Delta d \sim 100$ nm and frequency $\Omega = 256$ kHz, $d = d_0 + \Delta d \cos \Omega t$. Demodulation of the detector signal at a frequency $n\Omega$ ($n = 3$) then suppresses terms $|\mathbf{E}_A|^2$, $2\text{Re}\{\mathbf{E}^{\text{ref}*} \cdot \mathbf{E}_A\}$ and $|\mathbf{E}^{\text{ref}}|^2$ in Eq. (37). However, demodulation alone cannot suppress the interference of field-enhanced molecular scattering, $\mathbf{E}_{AOA}$, with the direct antenna scattering, $\mathbf{E}_A$ (term $2\text{Re}\{\mathbf{E}_A^* \cdot \mathbf{E}_{AOA}\}$ in Eq. (37)). To suppress this term and isolate $2\text{Re}\{\mathbf{E}^{\text{ref}*} \cdot \mathbf{E}_{AOA}\}$, it is further necessary to modulate the phase of $\mathbf{E}^{\text{ref}}$ by translating the reference arm mirror (analogous to s-SNOM where metal tips are used to measure the complex refractive index of a sample)[32]. Note that translation of the reference arm mirror also provides for spectral analysis of $\mathbf{E}_{AOA}$ following the Fourier-transform approach: (i) Recording of the demodulated detector signal as a function of the reference mirror position, (ii) Removal of the global phase offset to determine the sign of the interferogram signal (iii) Application of Tukey window with parameter $\alpha = 0.1$ (iv) Zero-filling by a factor of 3 (v) Performing a fast Fourier transform to obtain the raw near-field amplitude and phase spectra, $s_3^{\text{raw}}$ and $\varphi_3^{\text{raw}}$. In Extended Data Fig. 6(a) we show spectrally integrated near-field maps of the rod antennas. While the fundamental dipolar mode is observed on the rod antennas, as expected, importantly, the signal on the substrate is very small and below the noise floor. This was already observed for this setup in ref. [25]. Line profiles taken across the long axis of the rod quantity this observation (Extended Data Fig. 6(b)). Performing near-field spectroscopy on the antenna and on the substrate further reveals that the near-field signal is near zero on the substrate (Extended Data Fig. 6(c)). Therefore, we can conclude that (i) demodulation suppresses any direct contribution of the PDMS-layer and (ii) the vibrational features observed in Fig. 3 solely stem from the molecules (on the tip) scattering via the antenna after being illuminated by the antenna, i.e. the field-enhanced molecular scattering $\mathbf{E}_{AOA}$.

The raw near-field spectra of antennas were normalized to a spectrally flat reference. To this end, we fabricated a large (100 μm x 100 μm) metal (Au) patch that was located adjacent to the antennas on the same sample. We then carried out the following steps. First, the s-SNOM probe was moved to the center of the metal patch and approached to the patch surface. By doing so, the incident beam was focused on, and reflected at, the metal patch (the s-SNOM probe was shielded by the large metal patch and thus did not play a role). We then recorded the DC signal of the MCT detector (using the DC output at the detector amplifier and selecting the signal O0 in the microscope software) and acquired an interferogram. Applying the same steps as above yielded a DC amplitude and phase reference spectrum, $s_{\text{DC}}^{\text{ref}}(\omega)$ and $\varphi_{\text{DC}}^{\text{ref}}(\omega)$. The normalized near-field spectra, $\tilde{s}_3(\omega) = s_3(\omega)e^{i\,\varphi_3(\omega)}$, were then obtained with

$$s_3(\omega) = s_3^{\text{raw}}(\omega)/s_{\text{DC}}^{\text{ref}}(\omega) \text{ and } \varphi_3(\omega) = \varphi_3^{\text{raw}}(\omega) - \varphi_{\text{DC}}^{\text{ref}}(\omega) + 180°\,. \qquad (38)$$

Normalization of the amplitude spectra removed the nano-FTIR source spectrum. Normalization of the phase spectra provided an absolute measurement of the scattering phase of the antenna that corrects for the phase shifts accumulated by the illuminating and scattered beam when traversing the substrate, thus allowing for direct comparison of the phase spectra among different antennas. Note that in the reference measurement, the illuminating beams is subjected to a phase shift of 180° when reflecting from the metal patch, thus we explicitly added this phase shift to $\varphi_3$ in Eq. (38). To obtain near-field spectra over a large spectral range, we performed above experiment for different settings of the nano-FTIR laser (labeled settings B, C and D in the software, providing coverage from 850 to 1450 cm$^{-1}$, 1200 to 1900 cm$^{-1}$ and 1500 to 2200 cm$^{-1}$, respectively). We stitched together the individual spectra to a full spectrum by employing a smooth transition at wavenumber 1350 cm$^{-1}$ (between B and C) and 1640 cm$^{-1}$ (between C and D), following the protocol from ref.[33].

The error bars in Fig. 3(c,f) were obtained as follows. We estimated the measurement noise by calculating the root-mean-square of the complex-valued signal in the near-field spectra, $ns = \text{RMS}\,[s_3^{\text{raw}}(\omega)\exp i\varphi_3^{\text{raw}}(\omega)]$, in a range of 2,600 and 3,500 cm$^{-1}$, where no near-field signal is expected as it is outside the spectral range of the nano-FTIR source. It can be shown that the error in the normalized amplitude and phase signal is obtained by $\text{err}[s_3(\omega)] = ns/\left(\sqrt{2}s_{\text{DC}}^{\text{ref}}(\omega)\right)$ and $\text{err}[\varphi_3(\omega)] = ns/s_3^{\text{raw}}(\omega)$, respectively.

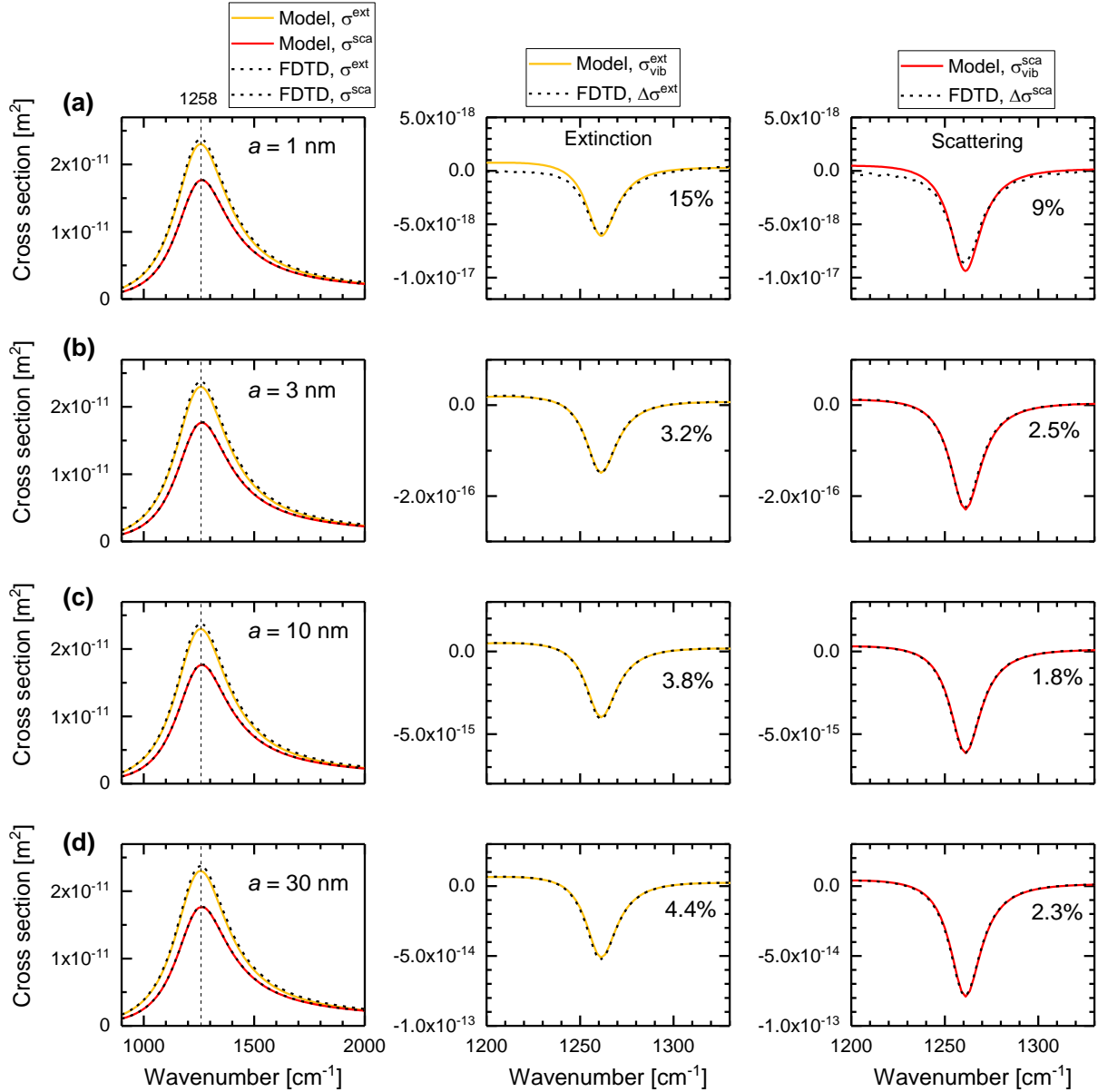

**Extended Data Figure 1: Numerical validation of the scattering model. Variation of particle radius.** (a)-(d) An IR-resonant metal nanorod of length $L = 3.19$ µm couples to a small particle with of radius $a = 1$ nm, 3 nm, 10 nm, 30 nm, respectively, and a vibrational resonance at 1,258 cm$^{-1}$. First column: Modeled (Eqs. (11),(12)) and numerically calculated extinction and scattering cross section, $\sigma^{\text{ext}}$ and $\sigma^{\text{sca}}$. Second and third columns: Spectral signature of the molecular vibration in the extinction and scattering cross section as obtained with the scattering model, $\sigma_{\text{vib}}^{\text{ext}}$ and $\sigma_{\text{vib}}^{\text{sca}}$ (Eq. (13)), compared to numerical calculations, $\Delta\sigma^{\text{ext}}$ and $\Delta\sigma^{\text{sca}}$. The number given in the plots is the maximum relative error between calculation and the scattering model. Note larger error observed with small particle diameter $a = 1$ nm is attributed to numerical error.

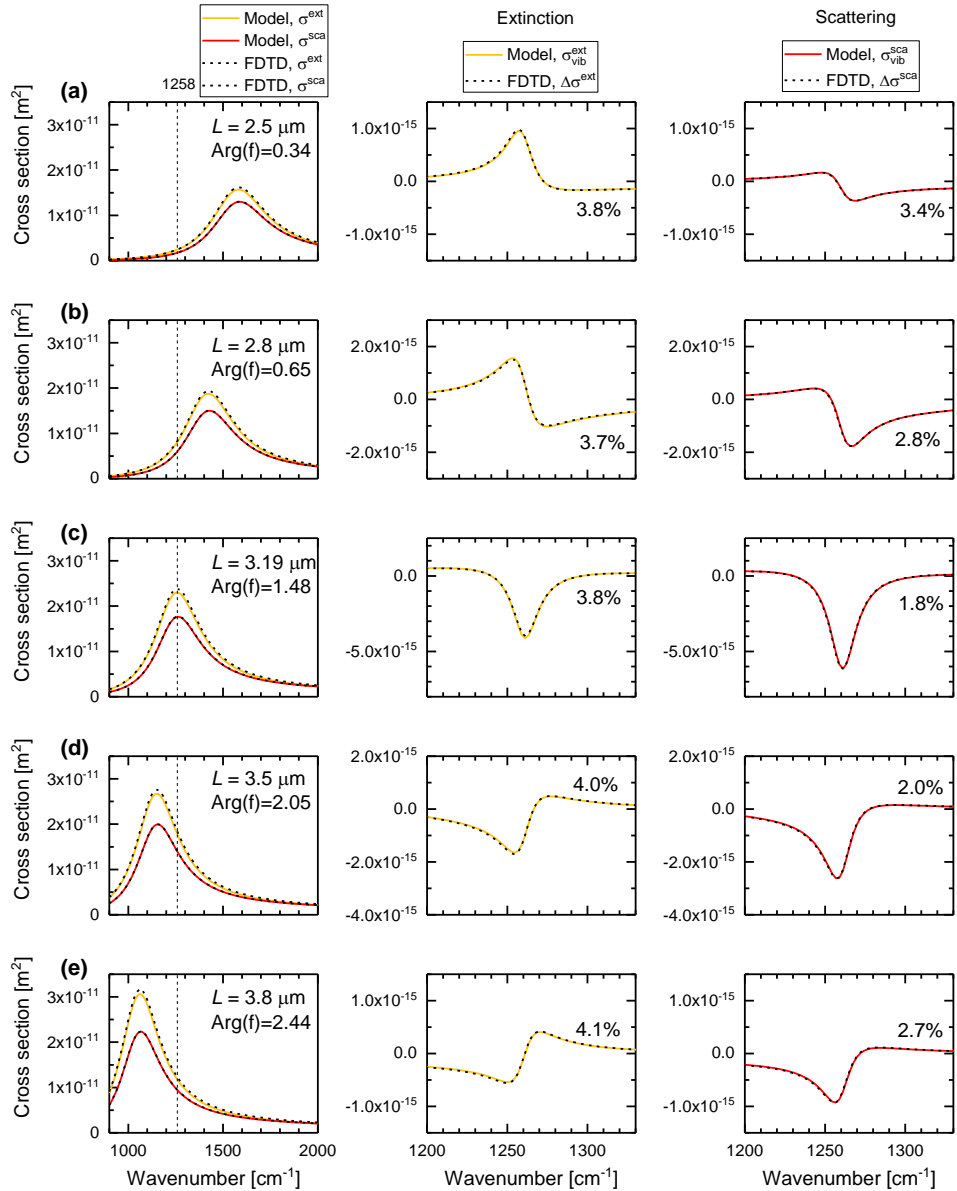

**Extended Data Figure 2: Numerical validation of the scattering model. Variation of antenna length.** (a)-(e) An IR-resonant metal nanorod of length $L = 2.5$ µm, $2.8$ µm, $3.19$ µm, $3.5$ µm, $3.8$ µm, respectively, couples to a small particle with 10 nm radius and a vibrational resonance at 1,258 cm$^{-1}$. First column: Modeled (Eqs. (11),(12)) and numerically calculated extinction and scattering cross section, $\sigma^{\text{ext}}$ and $\sigma^{\text{sca}}$. $\text{Arg}(f)$ specifies the phase (in radians) of the field enhancement $f$ provided by the antenna, evaluated at the vibrational resonance at 1,258 cm$^{-1}$, which controls the observed line shapes. Second and third columns: Spectral signature of the molecular vibration in extinction and scattering cross section as obtained with the scattering model, $\sigma_{\text{vib}}^{\text{ext}}$ and $\sigma_{\text{vib}}^{\text{sca}}$ (Eq. (13)), compared to numerical calculations, $\Delta\sigma^{\text{ext}}$ and $\Delta\sigma^{\text{sca}}$. The number given in the plots is the maximum relative error between calculation and the scattering model.

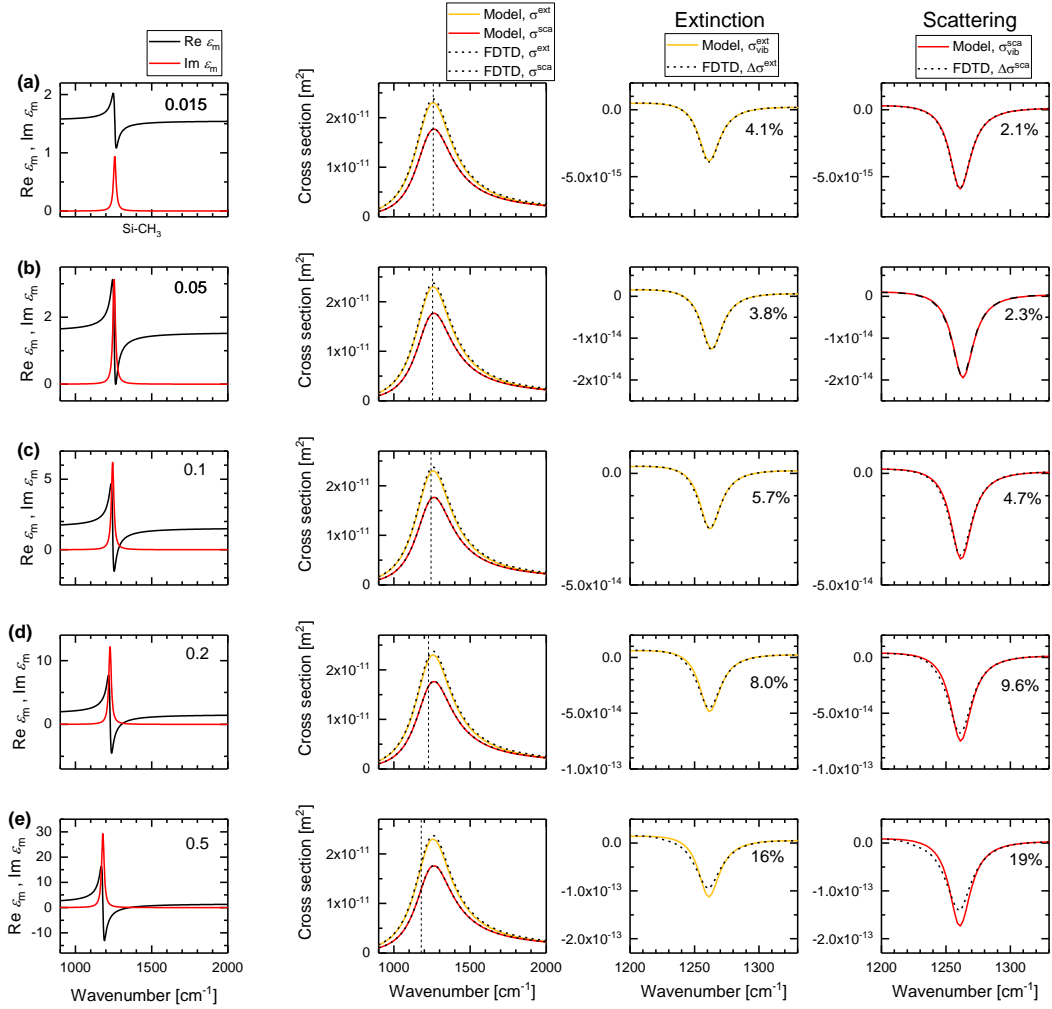

**Extended Data Figure 3: Numerical validation of the scattering model. Variation of oscillator strength.** (a-e) an IR-resonant metal antenna coupled to a small particle with 10 nm radius and a vibrational resonance modeled with different oscillator strengths: $\varepsilon_{\text{Lorentz}} = 0.015, 0.05, 0.1, 0.2, 0.5$ and $\omega_{\text{Lorentz}} = 1{,}258, 1{,}254, 1{,}244, 1{,}227$ and $1{,}170 \text{ cm}^{-1}$, respectively. First column: Object permittivity, $\varepsilon_O(\omega)$. Second column: Modeled (Eqs. (11),(12)) and numerically calculated extinction and scattering cross section, $\sigma^{\text{ext}}$ and $\sigma^{\text{sca}}$. Third and fourth columns: Spectral signature of the molecular vibration in extinction and scattering cross section as obtained with the scattering model, $\sigma_{\text{vib}}^{\text{ext}}$ and $\sigma_{\text{vib}}^{\text{sca}}$ (Eq. (13)), compared to numerical calculations, $\Delta\sigma^{\text{ext}}$ and $\Delta\sigma^{\text{sca}}$. The number given in the plots is the maximum relative error.

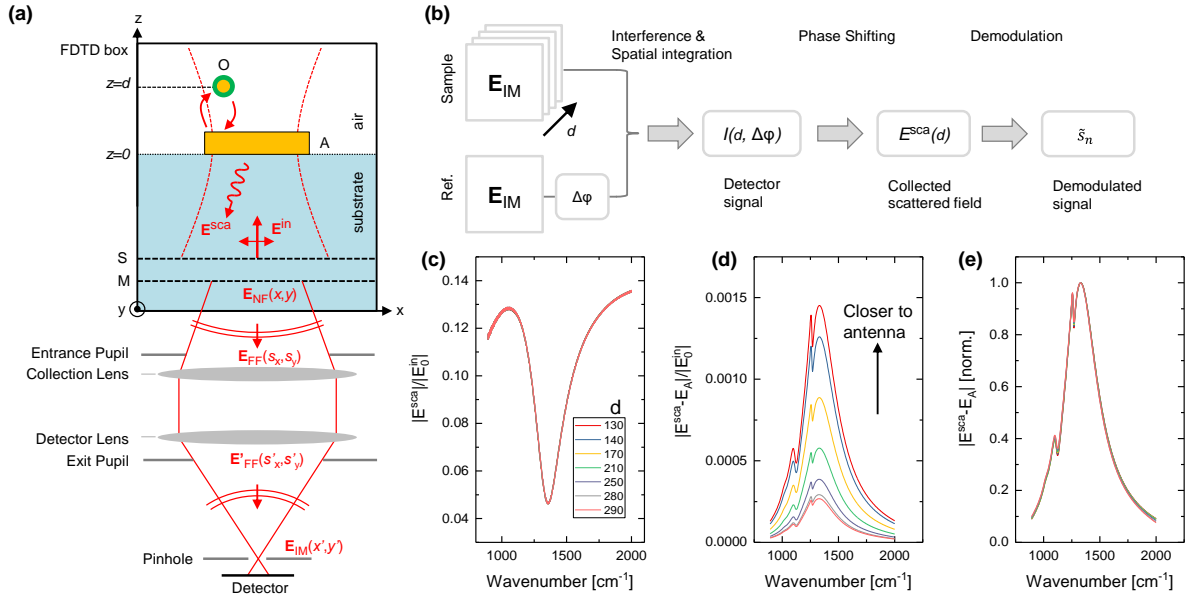

**Extended Data Figure 4: Numerical calculation of the near-field experiment.** (a) Numerical model which describes the near-field interaction between antenna (A) and tip (O) using the Finite-Difference Time-Domain (FDTD) method, but also takes into account the optical apparatus to accurately describe the light collection. (b) Illustration of the processing chain where the images obtained from the FDTD simulations, $E_{\mathrm{IM}}(x', y')$, are used to produce a numerically calculated near-field signal, $\tilde{s}_n = s_n e^{i\varphi_n}$, including the aspect of demodulation. (c-e) Illustration of the calculated fields before demodulation is applied. (c) Calculated reflected field, $E^{\mathrm{sca}}$, for the 2.5 μm long nanorod in Fig. 3 for a few selected values for the antenna-object distance, $d$. (d) Differential spectra, $E^{\mathrm{sca}} - E_A$. (e) Normalized differential spectra.

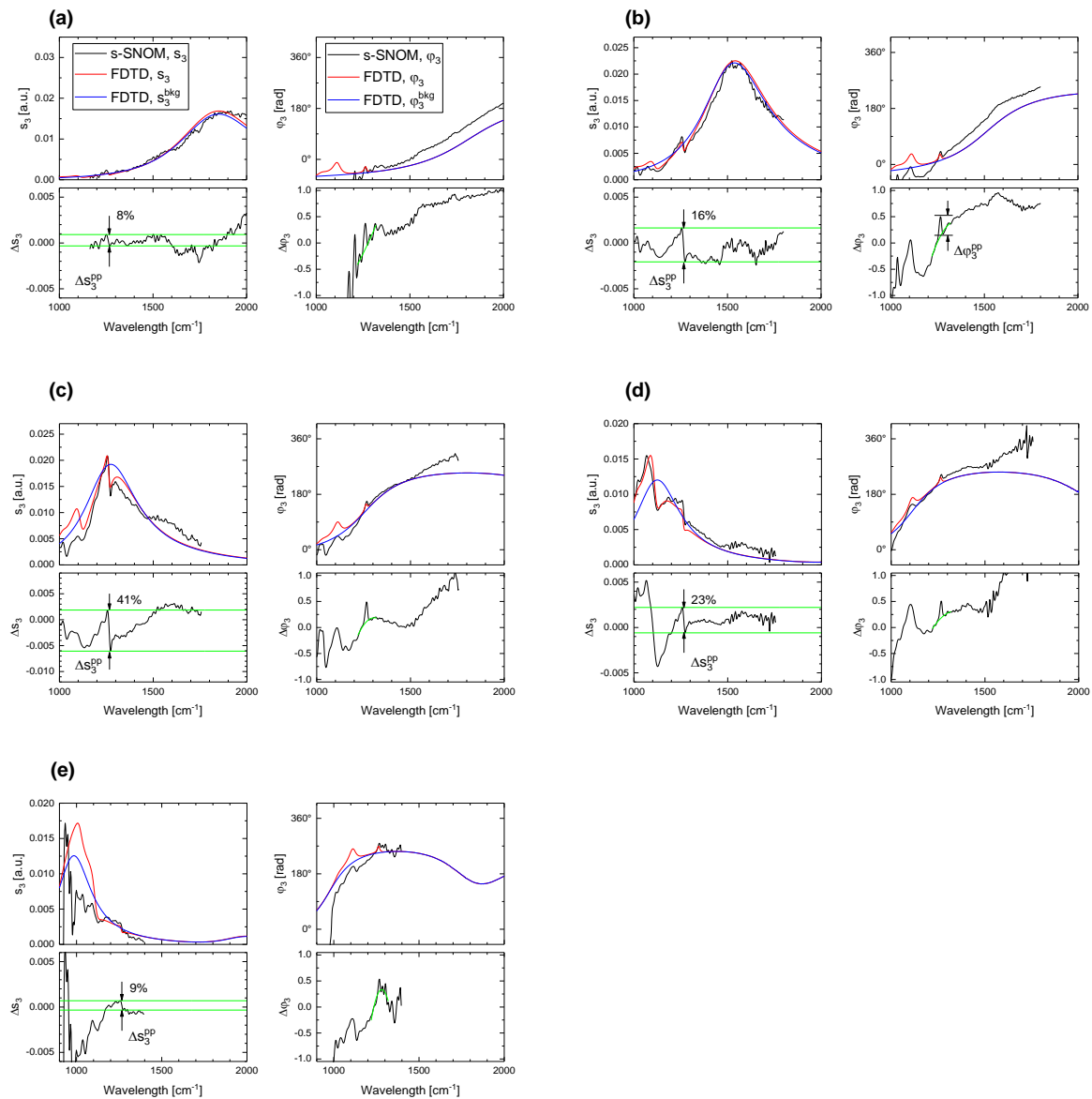

**Extended Data Figure 5: Extraction of the spectral signature of the molecular vibration for Fig. 3.** Spectra as obtained with the tip position at the left end of the nanorod. (a) Nanorod length, $L = 1.7$ μm, (b) $L = 2.1$ μm, (c) $L = 2.6$ μm, (d) $L = 3.0$ μm and (e) $L = 3.5$ μm. $s_3$ and $\varphi_3$: experimental and calculated amplitude and phase spectra, obtained with a PDMS-contaminated s-SNOM probe. $s_3^{bkg}$ and $\varphi_3^{bkg}$: numerically calculated amplitude and phase spectra assuming an Au nanoparticle covered by non-absorbing dielectric. Bottom left panels: values indicate the contrast of the spectral signature of the 1,258 cm$^{-1}$ (Si-CH$_3$) molecular vibration of PDMS relative to the maximum of the antenna scattering.

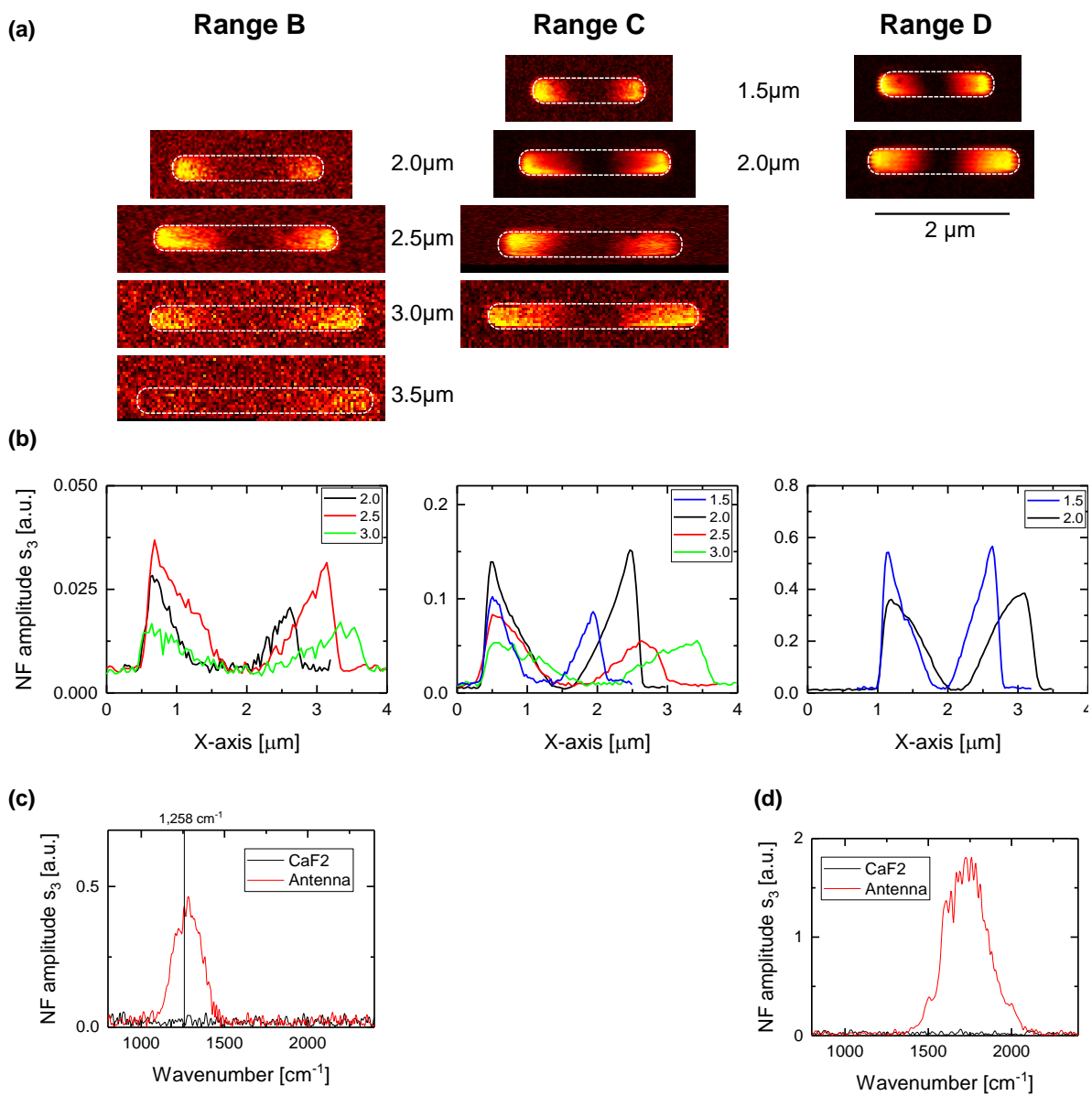

**Extended Data Figure 6: White-light images of the rod antennas and other test measurements.** (a) White-light images of the rod antennas taken at different settings of the nano-FTIR laser (B: 850 to 1450 cm$^{-1}$, C: 1200 to 1900 cm$^{-1}$ and D: 1500 to 2200 cm$^{-1}$, rod antenna length is stated next to the images). The fundamental dipolar mode is clearly visible. Note that signal-to-noise is limited compared to the near-field spectra shown in Extended Data Fig. 5 owing to the short integration time. (b) Line profiles taken along the long rod axis in (a), averaged over the width of the rods. (c,d) Raw near-field amplitude spectra obtained on the antenna and on the adjacent CaF$_2$ substrate. A 2.5 µm long rod antenna was measured in (c) and a 1.8 µm rod antenna in (d).

|  | Nanoparticle radius $a$ | 1 nm | 3 nm | 10 nm | 30 nm |
|---|---|---|---|---|---|
| Small particle | $\sigma_O^{sca}$ [$m^2$] | $2.3 \cdot 10^{-31}$ | $1.69 \cdot 10^{-28}$ | $2.32 \cdot 10^{-25}$ | $1.69 \cdot 10^{-22}$ |
| Small particle | $\sigma_O^{abs}$ [$m^2$] | $2.2 \cdot 10^{-21}$ | $6.00 \cdot 10^{-20}$ | $2.22 \cdot 10^{-18}$ | $6.00 \cdot 10^{-17}$ |
| Field-enhcd. | $|f|^4 \sigma_O^{sca}$ [$m^2$] | $1.7 \cdot 10^{-24}$ | $1.05 \cdot 10^{-21}$ | $7.58 \cdot 10^{-19}$ | $1.25 \cdot 10^{-16}$ |
| Field-enhcd. | $|f|^2 \sigma_O^{abs}$ [$m^2$] | $6.1 \cdot 10^{-18}$ | $1.49 \cdot 10^{-16}$ | $4.01 \cdot 10^{-15}$ | $5.15 \cdot 10^{-14}$ |
| Extinction | FDTD $\Delta\sigma^{ext}$ [$m^2$] | $-5.8 \cdot 10^{-18}$ | $-1.50 \cdot 10^{-16}$ | $-4.12 \cdot 10^{-15}$ | $-5.28 \cdot 10^{-14}$ |
| Extinction | Model $\sigma_{vib}^{ext}$ [$m^2$] | $-5.9 \cdot 10^{-18}$ | $-1.49 \cdot 10^{-15}$ | $-3.99 \cdot 10^{-15}$ | $-5.12 \cdot 10^{-12}$ |
| Scattering | FDTD $\Delta\sigma^{sca}$ [$m^2$] | $-8.6 \cdot 10^{-18}$ | $-2.26 \cdot 10^{-16}$ | $-6.16 \cdot 10^{-15}$ | $-7.89 \cdot 10^{-14}$ |
| Scattering | Model $\sigma_{vib}^{sca}$ [$m^2$] | $-9.2 \cdot 10^{-18}$ | $-2.29 \cdot 10^{-15}$ | $-6.15 \cdot 10^{-15}$ | $-7.91 \cdot 10^{-12}$ |
| Enh. Factors | $F^{ext}$ | $2.9 \cdot 10^{13}$ | $1.04 \cdot 10^{12}$ | $2.04 \cdot 10^{10}$ | $3.61 \cdot 10^{8}$ |
| Enh. Factors | $F^{sca}$ | $4.5 \cdot 10^{13}$ | $1.60 \cdot 10^{12}$ | $3.13 \cdot 10^{10}$ | $5.54 \cdot 10^{8}$ |

**Extended Data Table 1: Comparison between molecular scattering and molecular absorption.** *Small particle* scattering and absorption cross sections of a spherical nanoparticle with a vibrational resonance as obtained analytically with $\sigma_O^{sca} = \frac{k^4}{6\pi}|\alpha_O|^2$ and $\sigma_O^{abs} = k\,\text{Im}\{\alpha_O\}$, respectively, where $k$ is the magnitude of the free-space wavevector and $\alpha_O$ is the object polarizability. *Field-enhanced* scattering and absorption cross sections of the small particle, $|f|^4\sigma_O^{sca} = \frac{k^4}{6\pi}|f|^4|\alpha_O|^2$ and $|f|^2\sigma_O^{abs} = k|f|^2\text{Im}\,\alpha_O$. *Extinction & Scattering*: Numerically calculated (FDTD) spectral signature of the molecular vibration, $\Delta\sigma = \sigma - \sigma_{bkg}$, and modeled spectral signature, $\sigma_{vib}^{ext}$ and $\sigma_{vib}^{sca}$ (Eq. (13)), in the SEIRA extinction and scattering cross section. *Enhancement factors*: Combined enhancement provided by interferometric and field enhancement of molecular scattering in SEIRA extinction and scattering, $F^{ext}$ and $F^{sca}$ (Eqs. (15), (16)). All data are evaluated at the vibrational resonance at 1,258 cm$^{-1}$. The corresponding spectra are shown in Extended Data Fig. 1.